\documentclass[preprint,12pt]{elsarticle}

\usepackage{multirow}
\usepackage{tikz,ifthen,calc,graphicx,pgfplots,subfig}
\usepackage{amsmath,amsthm}
\usepackage{amsfonts,amssymb}
\usepackage{mathrsfs}
\usepackage{color,bm}
\usepackage[mode=buildnew]{standalone}

\newcounter{excercise}
\newcounter{excercisepart}

\definecolor{pennblue}{cmyk}{1,0.65,0,0.30}
\definecolor{pennred}{cmyk}{0,1,0.65,0.34}
\definecolor{mygreen}{rgb}{0.10,0.50,0.10}

\renewcommand \P     [1] {\text{\normalfont P}   \left[#1\right]}

\def \reals    {{\mathbb R}}

\def\ccalE{{\ensuremath{\mathcal E}}}
\def\ccalF{{\ensuremath{\mathcal F}}}
\def\ccalG{{\ensuremath{\mathcal G}}}

\def\ccalK{{\ensuremath{\mathcal K}}}

\def\ccalM{{\ensuremath{\mathcal M}}}
\def\ccalN{{\ensuremath{\mathcal N}}}

\def\ccalS{{\ensuremath{\mathcal S}}}

\def\ccal0{{\ensuremath{\mathcal 0}}}
\def\bbC{{\ensuremath{\mathbf C}}}
\def\bbD{{\ensuremath{\mathbf D}}}

\def\bbH{{\ensuremath{\mathbf H}}}

\def\bbL{{\ensuremath{\mathbf L}}}

\def\bbP{{\ensuremath{\mathbf P}}}

\def\bbW{{\ensuremath{\mathbf W}}}
\def\bbU{{\ensuremath{\mathbf U}}}
\def\bbV{{\ensuremath{\mathbf V}}}
\def\bbS{{\ensuremath{\mathbf S}}}

\def\bbZ{{\ensuremath{\mathbf Z}}}

\def\bbh{{\ensuremath{\mathbf h}}}

\def\bbv{{\ensuremath{\mathbf v}}}

\def\bbx{{\ensuremath{\mathbf x}}}
\def\bby{{\ensuremath{\mathbf y}}}

\def\bb0{{\ensuremath{\mathbf 0}}}
\def\hbZ{{\hat{\ensuremath{\mathbf Z}} }}
\def\hbh{{\hat{\ensuremath{\mathbf h}} }}

\def\hbx{{\hat{\ensuremath{\mathbf x}} }}
\def\hby{{\hat{\ensuremath{\mathbf y}} }}

\def\tby{{\tilde{\ensuremath{\mathbf y}} }}

\def\bbalpha{\boldsymbol{\alpha}}

\def\bbDelta{\boldsymbol{\Delta}}

\def\bbXi{\boldsymbol{\Xi}}

\usetikzlibrary{positioning}

\newtheorem{definition}{Definition}

\newtheorem{proposition}{Proposition}

\def\QED{{\setlength{\fboxsep}{0pt}\setlength{\fboxrule}{0.2pt}\fbox{\rule[0pt]{0pt}{1.3ex}\rule[0pt]{1.3ex}{0pt}}}}

\usepackage{subfig}

\def\B{{\mathbf B}}
\def\C{{\mathbf C}}
\def\D{{\mathbf D}}
\def\W{{\mathbf W}}

\def\P{{\mathbf P}}

\def\Z{{\mathbf Z}}
\def\I{{\mathbf I}}

\def\Q{{\mathbf Q}}

\def\U{{\mathbf U}}
\def\V{{\mathbf V}}

\def\miH{{\mathbf H}}
\def\S{{\mathbf S}}
\def\T{{\mathbf T}}

\def\w{{\mathbf w}}

\def\x{{\mathbf x}}

\def\y{{\mathbf y}}
\def\z{{\mathbf z}}

\def\h{{\mathbf h}}

\def\d{{\mathbf d}}

\def\t{{\mathbf t}}

\def\0{{\mathbf 0}}

\def\Lambdab{{\boldsymbol \Lambda}}
\def\Deltab{{\boldsymbol \Delta}}
\def\Psib{{\boldsymbol \Psi}}

\def\Thetab{{\boldsymbol \Theta}}

\newcommand{\new}[1] {{\textcolor{black}{#1}}}

\journal{Signal Processing}

\begin{document}

\begin{frontmatter}

\title{Graph-signal Reconstruction and Blind Deconvolution for Structured Inputs}

\author{David~Ram\'irez$^{1}$, Antonio~G.~Marques$^{2}$, and~Santiago~Segarra$^{3}$}
\address{$^{1}$ Department of Signal Theory and Communications, Universidad Carlos III de Madrid, Legan\'es, Spain and with the Gregorio Mara\~n\'on Health Research Institute, Madrid, Spain \\ 
          $^{2}$ Department of Signal Theory and Communications, King Juan Carlos University, Madrid, Spain \\ 
          $^{3}$ Department of Electrical and Computer Eng., Rice University, Houston, TX, USA } 

\begin{abstract}

Key to successfully deal with complex contemporary datasets is the development of tractable models that account for the irregular structure of the information at hand. 
\new{This paper provides a comprehensive and unifying view of several sampling, reconstruction, and recovery problems for signals defined on irregular domains that can be accurately represented by a graph.} 
The workhorse assumption is that the (partially) \textit{observed signals} can be \textit{modeled as the output of a graph filter} to a structured (parsimonious) input graph signal.  
When either the input or the filter coefficients are known, this is tantamount to assuming that the signals of interest live on a subspace defined by the supporting graph. 
When neither is known, the model becomes bilinear. Upon imposing different priors and additional structure on either the input or the filter coefficients, a broad range of relevant problem formulations arise. 
\new{The goal is then to leverage those priors, the shift operator of the supporting graph, and the samples of the signal of interest to recover: the signal at the non-sampled nodes (graph-signal interpolation), the input (deconvolution), the filter coefficients (system identification), or any combination thereof (blind deconvolution).} 
\end{abstract}


\end{frontmatter}

\section{Introduction}

Graph signal processing (GSP) generalizes traditional signal processing (SP) algorithms to deal with signals defined on irregular domains represented by graphs \cite{Shuman2013,Sandryhaila2014,GSP_overview2017}. Depending on the application, the particular graph may correspond to an actual (social, electrical, sensor) network where the signal is observed, or encode (pairwise) statistical relationships between the signal values. \new{Recent examples of GSP works dealing with relevant problems include sampling and reconstruction of signals \cite{Chen2014,Chamon2018,Sakiyama2019,Tanaka2020_spm}, with an early focus on graph bandlimited models \cite{SamplingOrtegaICASSP14,Kovacevic2015,Marques2016Sampling,Tsitsvero2015signals, Richard_Djuric_book,DiLorenzo2018}, filter design \cite{Loukas2015,segarra_2017_optimal}, frequency analysis \cite{Sandryhaila2014a,Marques2016}, and blind deconvolution \cite{Segarra2016,iwata2020graph}, to name a few.} 

This paper provides a comprehensive \new{and unifying} view of recovery and reconstruction problems involving graph signals. The common denominator throughout all of them is the assumption that the linear model $\bby=\bbH\bbx$ holds, where $\bby$ is an \textit{observed} graph signal, $\bbH$ is a \textit{linear graph filter}, and $\bbx$ is an unknown \textit{structured} input. Building on this model and assuming that we have access to a) the values of $\bby$ at a subset of nodes, b) the \new{graph shift operator}, and c) side information on $\bbH$ and $\bbx$, the goal is to recover i) the values of $\bby$ at the non-observed nodes (interpolation), ii) the graph filter $\bbH$ (system identification), iii) the values or support of $\bbx$ (deconvolution), or iv) any combination of the previous ones (e.g., blind deconvolution). \new{These span a broad range of (inverse) graph-signal reconstruction problems, introducing new and existing formulations within a common framework, generalizing various results available in the literature, and developing new algorithms.} Moreover, since graph filters can be efficiently used to model local diffusion dynamics~\cite{segarra_2017_optimal,Segarra2016}, the relevance of the developed schemes goes beyond signal reconstruction, being particularly pertinent in applications such as opinion formation and source identification in social networks, inverse problems of biological signals supported on graphs, and modeling and estimation of diffusion processes in multi-agent networks. Our main interest is in setups where the input graph signal $\bbx$ is sparse, with the associated output signal $\bby$ in that case being referred to as a \textit{diffused sparse graph signal}. Nonetheless, the paper also generalizes the results obtained for diffused sparse graph signals to setups where $\bbx$ belongs to a subspace. 

\subsection*{Outline, contributions, and related work}

\begin{figure}[t]
     \centering
     \includegraphics[width=0.85\textwidth]{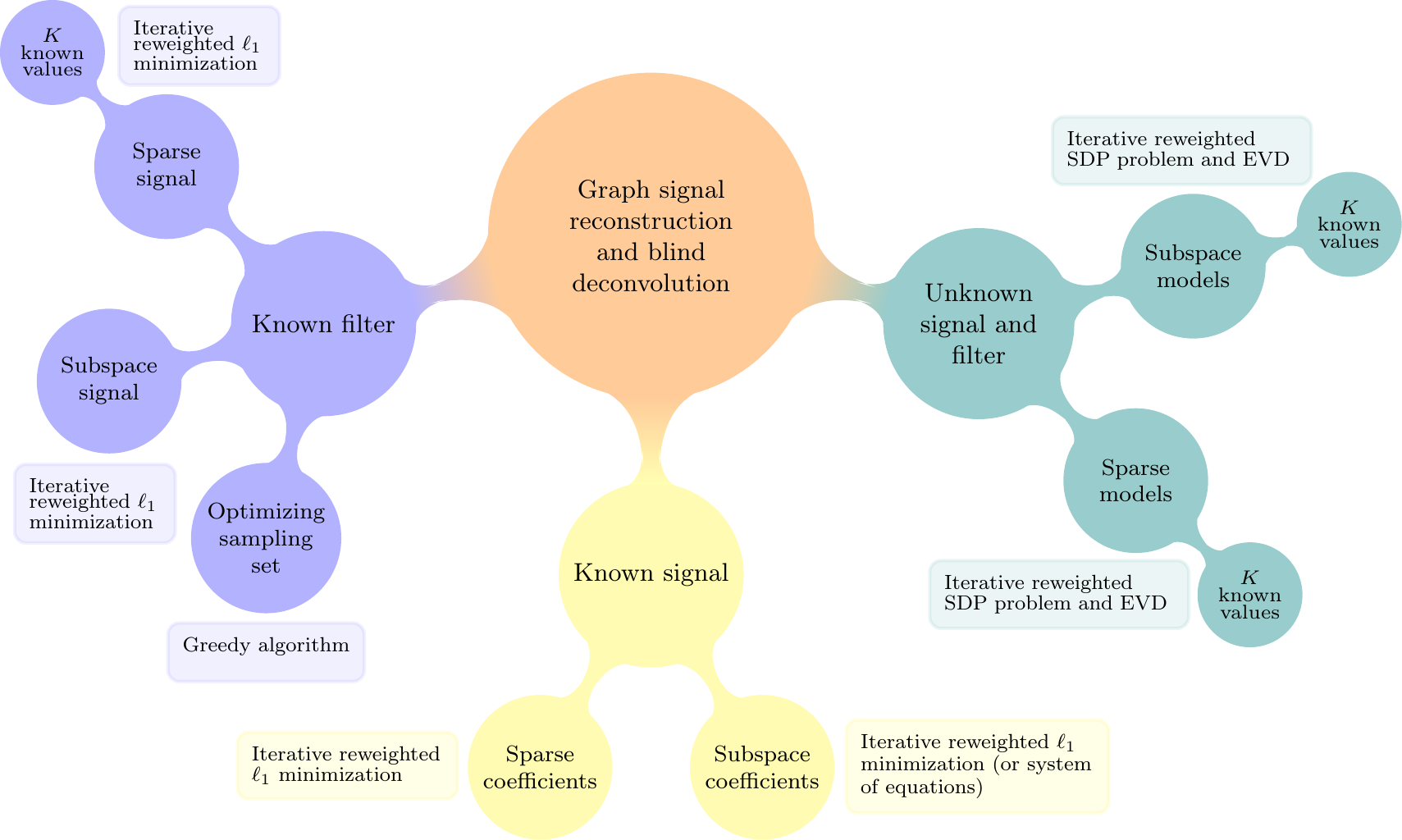}
     \caption{\new{Signal reconstruction and blind deconvolution problems studied in this work, including the approach to solve them. In the figure, SDP and EVD stand for semidefinite programming and eigenvalue decomposition, respectively.}}
     \label{fig:diagram}
\end{figure}

We first look at the more favorable setup where the filter $\bbH$ is known (Section \ref{S:Recovery_Known_Filters}). The goal there is to interpolate $\bby$ from a few nodal samples under the assumption that the input $\bbx$ is sparse. The support of $\bbx$ is not known and additional information on some of the values of $\bbx$ may be available or not. \new{This problem falls into the class of sparse signal reconstruction and compressive sensing \cite{Baraniuk2007,Ranieri2014,Elad2007,Learning_sense_Duarte}.} While a number of GSP works have investigated the reconstruction of bandlimited signals (i.e., assuming that $\bby$ belongs to a subspace defined by some of the frequencies of the graph \cite{SamplingOrtegaICASSP14,Kovacevic2015,Marques2016Sampling,Tsitsvero2015signals}), the \textit{subspace} here is given by the columns of filter $\bbH$. Moreover, our focus is on \textit{blind} setups where the support of the input is not known. Hence, the results are also relevant in the context of source identification. 
\new{In the second investigated setup, the input $\bbx$ and the graph shift operator are known, but the filter coefficients are not (Section \ref{sec:filter_recovery}). The goal then is to use the sampled version of the signal $\bby$ to reconstruct the signal at the unobserved nodes as well as the filter coefficients. While filter identification algorithms using input-output pairs exist \cite{segarra_2017_optimal}, we focus on graph signal reconstruction \cite{Tanaka2020_spm} and incorporate side information.} We finally transition to setups where neither the coefficients that define the filter $\bbH$ nor the input signal $\bbx$ are known (Section~\ref{S:jointblind}). 
In this case, we assume that the maximum degree of the filter is known and that some of the values of the input $\bbx$ may be available. The problem of joint filter and input identification for graph signals was first addressed in \cite{Segarra2016}. The difference here is on the \textit{algorithmic approach} (which yields better results), the incorporation of additional \textit{side information} on the input $\bbx$, and the interest in reconstructing $\bby$. Moreover, we do not only focus on the case where the input signals and filters are assumed to be sparse but also the case where they lie on a known subspace. \new{All in all, we provide a unifying framework for a number of graph signal reconstruction problems adhering to the model $\bby=\bbH\bbx$, address estimation setups not considered before (including the interpolation of $\bby$ with $\bbx$ known, the optimization of the sampling set, the identification of the graph filter coefficients under sparsity and subspace constraints, and the consideration of certain types of side information), and introduce new algorithms for their solution. A diagram summarizing the considered setups, along with the algorithms to address them, is depicted in Fig. \ref{fig:diagram}.}


\section{Preliminaries}
\label{sec:preliminaries}

This section briefly reviews GSP concepts used in the manuscript, presents the definition of diffused sparse graph signals ---which are the main focus of this work---, and discusses alternative subspace models for graph signals. 

We denote by $\mathcal{G} = (\ccalN,\ccalE)$ a directed graph composed by a set $\ccalN$ of $N$ nodes and a set of (possibly weighted) links $\mathcal{E}$. When there exists a link from node $i$ to node $j$, the element $(i,j)$ belongs to $\ccalE$. Given the graph $\ccalG$, we define a graph signal as a mapping from $\ccalN$ to $\reals$, which is conveniently represented by a vector $\x = [x_1, \ldots, x_N]^T \in \mathbb{R}^{N}$, with $x_i$ being the signal value at node $i$. To exploit the graph structure in the processing of graph signals, we define the shift operator $\S$ \cite{Sandryhaila2013,Sandryhaila2014a}. This is a sparse matrix whose sparsity pattern is given by $\mathcal{G}$ where $[\S]_{ji} \neq 0$ for $(i,j) \in \mathcal{E}$ or $i = j$. Typically, the graph shift operator is given by the adjacency matrix \cite{Sandryhaila2013,Sandryhaila2014a} or the graph Laplacian \cite{Shuman2013}, yet many other alternatives are possible. Interestingly, due to the structure of $\S$, the operation $y_i = [\S \x]_i$ may be computed locally, i.e., only the signal values $\{x_j\}_{j \in \mathcal{N}_i}$ are required, where $\mathcal{N}_i = \{j \, | \, (j,i) \in \mathcal{E}\}$. In this work, we assume that $\S$ is diagonalizable, which allows us to decompose it as $\S = \V \Lambdab \V^{-1}$, where $\Lambdab = \text{diag}(\lambda_1, \ldots, \lambda_N)$, with $\lambda_1 \geq \cdots \geq \lambda_N$.

Equipped with these definitions, we may introduce the concept of graph filters, which are simply graph-signal operators defined as polynomials in the graph shift \cite{Sandryhaila2013}, that is, $  \miH = \sum_{l = 0}^{L-1} h_l \S^l.$
Hence, the filtering becomes $\y = \miH \x$, where $\y$ and $\x$ are, respectively, the filtered and input signals, and $\bbh=[h_0, \ldots, h_{L-1}]^T$ are the filter coefficients, with $L - 1$ being the filter order.

\new{Alternatively, as in conventional SP, graph filters and signals possess a frequency (or Fourier) representation. Specifically, the graph Fourier operator for signals is defined as $\U = \V^{-1}$, which yields the graph Fourier transform (GFT) of the signal $\x$ given by $\tilde{\x} = \U \x$. On the other hand, the graph Fourier transform for filters is obtained as $\tilde{\h} = \Psib \h$, where $\Psib$ is an $N \times L$ Vandermonde matrix whose elements are $[\Psib]_{i,i'} = \lambda_{i}^{i'-1}$ \cite{segarra_2017_optimal}. Notice that while in classical discrete time processing we have that $\U=\Psib$ and $[\U]_{i,i'}=e^{-j\frac{2\pi}{N}(i-1)(i'-1)}$, this is not the case in GSP.} Additionally, there is an analog of the convolution theorem, given by \cite{Sandryhaila2014a} 
\begin{equation}
  \label{eq:filtering_frequency}
  \tilde{\y} = \U \y = \text{diag}\big(\Psib \h\big) \U \x =  \text{diag}\big(\tilde{\h}\big) \tilde{\x} = \tilde{\h} \circ \tilde{\x},
\end{equation}
with $\circ$ denoting the Hadamard (or element-wise) product. \new{After some mathematical manipulations~\cite{Segarra2016}, it can be shown that an equivalent representation of \eqref{eq:filtering_frequency}, to be used later on, is}
\begin{equation}
\label{eq:filtering_frequency_KatriRhao}
\tilde{\y} =  (\Psib^T \odot \U^T )^T \text{vec}(\x \h^T),
\end{equation}
where $\odot$ is the Khatri-Rao (or columnwise Kronecker) product and $\text{vec}(\cdot)$ is the vectorization operator, defined by stacking the columns of the argument.

\subsection{Diffused sparse graph signals}
\label{Ss:generation_diffused_sparse_inputs}

\begin{definition}\label{eqn:diffused_sparse_inputs}
Let $\miH \in \reals^{N\times N}$ be a graph filter, then the signal $\bby \in \reals^N$ is called a diffused sparse graph signal of order $S_x$ if it holds that $\bby=\bbH\bbx$, with $\|\bbx\|_0=S_x.$
\end{definition}
From the definition above it becomes clear that $\bby$ lies in a subspace of dimension (at most) $S_x$ spanned by a subset of the columns of $\bbH$. This subspace is determined by the filter coefficients, the support of $\x$, and the network topology represented by $\S$. To better understand diffused sparse graph signals, we define the $l$th shifted version of the input as $\bbx^{[l]}=\bbS\bbx^{[l-1]}$ and rewrite the filter output as $\bby = \sum_{l=0}^{L-1}h_l\bbx^{[l]},$
where 
\begin{equation}
  \label{E:FilterAsLocalExchanges_2}
x_i^{[l]} = [\S]_{ii} x_{i}^{[l-1]} + \sum_{j\in \ccalN_i} [\S]_{ij} x_{j}^{[l-1]}.
\end{equation}
That is, the filter output becomes a linear combination of shifted inputs, which are computed locally. Thus, defining $\bbx^{[0]}=\bbx$, it is possible to interpret the filtered signal $\bby$ as a steady-state signal generated by $\bbx^{[0]}$ after being \textit{diffused locally} by means of the successive application of the network dynamics captured by $\bbS$. 

Based on this interpretation, at the initial state, only a few nodes have a non-zero value since $\bbx^{[0]}=\bbx$, with $\x$ sparse. 
After the application of a given shift, i.e., the $l$th one, the information is spread (diffused) across the one-hop neighborhood of the nodes in the (non-zero) support of $\bbx^{[l-1]}$, as shown in \eqref{E:FilterAsLocalExchanges_2}. For filter orders larger than the diameter of the graph, the seeding values in $\bbx^{[0]}$ will have percolated across the entire network, rendering the model relevant to tackle real-world problems. These applications range from social networks where a rumor originated by a small group of people is spread across the network via local opinion exchanges to brain networks where an epileptic seizure emanating from few regions is later diffused across the brain~\cite{kramer2008_emergent, mathur2020_graph}.

\subsection{Alternative subspace models for graph signals}\label{Ss:alternative_subspace_signal_models}

Subspace models are widely used in SP, playing critical roles in tasks such as compression, denoising, and solving ill-posed inverse problems. 
When particularized to graph signals, the key question is how to postulate subspace models that are practically relevant, while accounting in a meaningful way for the structure of the supporting graph. 
Early works focused on signals that were bandlimited in the graph shift matrix $\bbS$. That is, signals $\bbx$ that can be written as $\bbx=\sum_{k \in \ccalF} \tilde{x}_k \bbv_k$ with $\ccalF=\{f_1, \ldots, f_{S_{\tilde{x}}}\}$ being the set of active frequencies, ${S_{\tilde{x}}}\ll N$ and $\bbv_k$ being the $k$th column of $\bbV=[\bbv_1,\ldots,\bbv_N]$. The motivation for considering such a graph bandlimited model was twofold. When $\bbS$ is set to the adjacency matrix of the supporting graph, this model is the natural generalization of bandlimited time-varying signals\footnote{If $\bbS$ is set to the adjacency of the directed cycle graph, which is the support of discrete time-varying periodic signals, then the eigenvectors $\bbV$ of the shift form the standard discrete Fourier transform matrix.} to more irregular graph domains \cite{Kovacevic2015,Marques2016Sampling}. When $\bbS=\bbL$ and $\{\bbv_k\}_{k \in \ccalF}$ correspond to the eigenvectors of the Laplacian matrix (those associated with the smallest eigenvalues), assuming bandlimitedness promotes reconstructed signals that are smooth across the graph. 

Since bandlimited graph signals have sparse representations in the frequency domain given by the GFT, a way to identify additional subspace models is to look at alternative (linear) transformations that had been shown to be meaningful for graph signals, with wavelets being one of the most prominent examples. To be more concrete, suppose that we are given a transform $\bbXi_x \in \reals^{N\times N}$ such that, given a graph signal $\bbx$, the transformed coefficients are computed as $\bbalpha_x = \bbXi_x \bbx$. It readily follows that if $\bbXi_x$ is invertible, one has that $\bbx = \bbXi^{-1}_x \bbalpha_x$. Since the goal is to design transforms $\bbXi$ for which $\bbalpha_x$ is sparse, this implies that the signal $\bbx$ lives on a subspace spanned by a subset of the columns of $\bbXi^{-1}_x$. A number of such transforms have been proposed in the literature, with relevant examples including the windowed graph Fourier transform \cite{shuman2016vertex}, diffusion wavelets \cite{coifman2006diffusion}, and spectral graph wavelets \cite{hammond2011wavelets}.

A more general approach is to consider dictionaries tailored to graph signals. To be more specific, consider a non-square dictionary $\bbD_x\in \reals^{N\times {D_x}}$ whose ${D_x}$ columns (atoms) represent graph signals. The idea is leveraging $\bbD_x$ to write the graph signal of interest as $\bbx=\bbD_x \bbalpha_x$ with $\bbalpha_x$ being a sparse vector of coefficients whose support indicates the atoms of the dictionary that are active. Different ways to build those dictionaries exist, including cluster-wise constant approaches, hierarchical schemes, and cases where $\bbD_x$ is learned from the data itself~\cite{yankelevsky2016dual}. 


Clearly, the subspace model proposed in Definition \ref{eqn:diffused_sparse_inputs} can be viewed as an alternative to any of the ones described in the paragraphs above. Equally interesting, one can combine the concepts in Sections \ref{Ss:generation_diffused_sparse_inputs} and \ref{Ss:alternative_subspace_signal_models}
 by applying the models in Section \ref{Ss:alternative_subspace_signal_models} to the input $\bbx$, which is later diffused as described in Section \ref{Ss:generation_diffused_sparse_inputs} to generate the graph signal of interest $\bby$. 
 This will be studied in more detail in the following sections.

\section{Recovery with known diffusing filters}
\label{S:Recovery_Known_Filters}

In this section, we investigate the recovery of a diffused sparse graph signal $\bby$ for the simplified setup where both the filter coefficients $\bbh$ and the graph shift $\bbS$ are known. Under these assumptions, the filtering matrix $\miH$ is completely known. The goal is then to use observations of $\y$ to either recover the values of $\bby$ in the non-observed nodes (interpolation), obtain the seeding values in $\bbx$ (network input estimation), identify the support of $\bbx$ (localization), or any combination thereof.

To precisely formulate this problem, we first define the \textit{sampling matrix} $\bbC_{\ccalM}\in \{0,1\}^{M\times N}$ whose rows correspond to canonical vectors identifying the elements $\ccalM=\{i_1, \ldots ,i_M\}$ of the signal $\bby$ that are observed. Based on this definition, the values at the observed nodes are $\bby_{\ccalM} = [y_{i_1}, \ldots, y_{i_M} ]^T = \bbC_{\ccalM}\bby = \bbH_{\ccalM} \bbx,$
with $\bbH_{\ccalM} = \bbC_{\ccalM} \bbH \in \mathbb{R}^{M  \times N}$ being the corresponding $M$ rows of $\miH$. Accordingly, the value of the graph signal at the unobserved nodes will be denoted as $\bby_{\ccalM^c} = \bbC_{\ccalM^c}\bby$, where $\ccalM^c=\ccalN \setminus \ccalM$. With this notation at hand, the recovery problem may therefore be formulated as
\begin{align}
\label{eq:minimization_l0}
\hbx= \mathop{\text{find}} \,\,  \{\bbx\}, \quad 
\text{s. to} \,\,\,  \bby_{\ccalM} = \bbH_{\ccalM} \x, \,\,\,  \|\x\|_{0}\leq S_x
\end{align}
from where we can then obtain $\hby_{\ccalM^c} = \bbC_{\ccalM^c}\bbH \hbx$. As can be seen from \eqref{eq:minimization_l0}, the recovery problem with known diffusing filter and a diffused sparse input signal is a classical problem in sparse signal reconstruction and compressive sensing \cite{Baraniuk2007}, for which we briefly summarize the recoverability conditions.

Before proceeding, we must also define the matrix $\bbC_{\ccalS_x} \in \{0,1\}^{S_x \times N}$ that encodes the sparsity in $\x$. That is, considering the set $\mathcal{S}_x = \{j_1,\ldots , j_{S_x}\}$, which contains the indices of the (unknown) support of $\x$, the matrix $\bbC_{\ccalS_x}$ yields the non-zero values associated with the $S_x$ nodes in $\mathcal{S}_x$. In the case of known support, the recoverability depends on the \textit{rank} (invertibility) of the submatrix $\bbC_{\ccalM} \bbH \bbC_{\ccalS_x}^T=\bbH_{\ccalM,\ccalS_x} \in \reals^{M\times S_x}$, which should be at least $S_x$ and, hence, requires $M \geq S_x$ as a necessary condition. For unknown supports, the recovery performance depends on the \textit{spark} of the matrix $\miH_{\ccalM}$. Concretely, in \cite{Marques2016Sampling} it was proved that for $M\geq 2S_x$, the optimization problem \eqref{eq:minimization_l0} provides perfect recovery if $\miH_{\ccalM}$ is full spark. Nonetheless, since \eqref{eq:minimization_l0} is not convex, the typical approach is to replace the $\ell_0$ norm with the convex $\ell_1$ norm \cite{Candes2008}. In this case, the recovery depends on the \textit{coherence} of the matrix $\miH_{\ccalM}$ \cite{Candes2008,Marques2016Sampling}. 

As mentioned above, the $\ell_1$ surrogate of the $\ell_0$ norm is typically used to solve the minimization form of the feasibility problem in \eqref{eq:minimization_l0}, that is,
\begin{equation}
\label{eq:minimization_l0_bis}
\hbx = \mathop{\text{argmin}}_{\x}  \,\, \|\x\|_1, \quad \text{s. to} \,\,\,  \bby_{\ccalM} = \bbH_{\ccalM} \x. 
\end{equation}
In some scenarios, for instance for small $M$, the performance of the $\ell_1$ norm surrogate is not sufficient, and a (typically) better alternative \cite{Candes2008} is $\|\x\|_0 \approx \sum_{n=1}^N \log(|x_n| + \epsilon_0)$, with $\epsilon_0$ being a small positive constant. The logarithm is a concave function, which yields a non-convex optimization problem. The solution to the optimization problem may be found using the majorization-minimization (MM) approach \cite{Palomar_MM}. The MM approach is an iterative technique composed of two steps. 
In the first one, a function that majorizes the cost and that is easy to optimize is sought and, in the second step, the majorizing function is minimized. These two steps are repeated until convergence. For the logarithm, a simple majorizing function is the first-order Taylor approximation. Then, with $i=1,\ldots,I,$ being an iteration index, the resulting minimization problem is
\begin{align}
\label{eq:minimization_reweighted_l11}
\hbx^{(i)} =\mathop{\text{argmin}}_{\x} \,\,  \sum_{n = 1}^{N} a_n |x_n|, 
\quad \text{s. to} \,\,\,  \bby_{\ccalM} = \bbH_{\ccalM} \x,
\end{align}
where $a_n = (|\hat{x}_n^{(i-1)}| + \epsilon_0)^{-1}$. Note that if we would define $a_n = 1$, \eqref{eq:minimization_reweighted_l11} simplifies to the $\ell_1$ norm minimization problem and just $I = 1$ iteration is necessary. 

We finally set the estimated input to $\hbx=\hbx^{(I)}$ and the interpolated output to $\hby=\bbH\hbx$. When there is noise in the observations and small model mismatches, we should substitute the linear constraint in \eqref{eq:minimization_reweighted_l11} with $\|\bby_{\ccalM} - \bbH_{\ccalM} \bbx\|_2^2 \leq \varepsilon$. Alternatively, if our main focus is on recovering the input $\bbx$ (rather that reconstructing $\bby$), the previous equation can be left-multiplied by the pseudo-inverse of $\bbH_{\ccalM}$. 

\subsection{Input graph signals lying on a subspace} 
\label{sec:subspace_signal}

The formulation just described can be easily modified to account for \textit{input signals} that belong to a subspace. To be specific, consider the signal model $\x = \bbD_x \bbalpha_x,$
where $\bbD_x$ may correspond to any of the wavelet or dictionary spaces presented in Section \ref{Ss:alternative_subspace_signal_models} and $\bbalpha_x$ is the sparse vector selecting the active columns. Under this subspace assumption for the input, we just need to replace $\bbx$ with $\bbalpha_x$ and $\bbH$ with $\bbH\bbD_x$ in \eqref{eq:minimization_l0}-\eqref{eq:minimization_reweighted_l11}. Once $\hat{\bbalpha}_x$ is found, the estimates for $\bbx$ and $\bby$ are simply obtained as $\hbx = \bbD_x\hat{\bbalpha}_x$  and $\hby_{\ccalM^c} = \bbC_{\ccalM^c}\bbH\bbD_x\hat{\bbalpha}_x$.

\subsection{Known input values}

We now study the reconstruction problem in \eqref{eq:minimization_reweighted_l11} when there is \emph{a priori} information on the input. In particular, we consider that $K$ values of the input are known. This knowledge can be based, for instance, on structural properties of the application (e.g., physiological constraints for particular nodes of a brain network), or, in multi-agent networks, it is reasonable that the sampling nodes could have access not only to the value of the diffused signal $\bby$, but also to their own value of the initial sparse input $\bbx$. 

The set of nodes where the signal values are known is $\mathcal{K} = \{k_1, k_2, \ldots , k_K\}$ and these values are $\bbx_{\ccalK} = \begin{bmatrix} x_{k_1}, \ldots, x_{k_K} \end{bmatrix}^T =\bbC_{\ccalK} \bbx$, with $\bbC_{\ccalK}\in \{0,1\}^{K \times N}$ being a binary matrix that selects the corresponding indices. For later use, we define $\ccalK^c = \mathcal{N} \setminus \mathcal{K}$ as the complement set of $\ccalK$, $\bbx_{\ccalK^c} = \bbC_{\ccalK^c} \x$ as the vector containing the $N - K$ unknown values of $\x$, with $\bbC_{\ccalK^c} \in \{0,1\}^{(N-K) \times N}$ the corresponding selection matrix. 
This knowledge can be easily incorporated into~\eqref{eq:minimization_reweighted_l11} by either augmenting the problem with the constraint $\bbx_{\ccalK}=\bbC_{\ccalK}\bbx$ or (more efficiently) by replacing the optimization variable $\bbx$ with $\bbx_{\ccalK^c}$, matrix $\bbH_{\ccalM}$ with $\bbH_{\ccalM} \bbC_{\ccalK^c}^T$, and the observations $\y_{\ccalM}$ with $\y_{\ccalM} - \bbH_{\ccalM} \bbC_{\ccalK}^T\bbx_{\ccalK}$.

\subsection{Optimizing the sampling set}
\label{Sec:optimizing_sampling}

So far, we have considered that the set of nodes $\mathcal{M}$ where the signal is sampled is given. Nonetheless, in some applications it is feasible to select the $M$ nodes that form $\mathcal{M}$. This section briefly reviews the challenges associated with the optimization of the sampling set and presents a simple algorithm for the setup at hand.

To find the optimal $\mathcal{M}$, one first requires i) a closed-form expression for the optimal signal estimator for a given sampling set and ii) an expression quantifying the error incurred by the optimal estimator in step i). If these steps are feasible, the resultant problem is well posed and, although typically NP-hard, different approximated algorithms, including greedy schemes, tend to perform well in practice \new{(See \cite{Ranieri2014} and references therein.)} However, for setups like the one in this paper, where \textit{the support} of the input \textit{is unknown}, not even step i) can be implemented, so that it is not possible to derive, to the best of the our knowledge, an expression quantifying the error as a function of $\mathcal{M}$.

\new{Given these challenges, and inspired by works in the area of sensor placement in sensor networks \cite{Ranieri2014} and projection optimization in compressed sensing \cite{Elad2007,Learning_sense_Duarte}, we shall use the following measure to optimize the sampling set}
\begin{equation}
  \label{eq:measure_orthogonality}
  \rho_{\mathcal{M}} = \frac{1}{M(M-1)} \sum_{i = 1}^{M} \sum_{\substack{j = 1 \\ j > i}}^{M} \frac{q_{ij}^2}{q_{ii} q_{jj}}, 
 \end{equation}
where $q_{ij}$ is the $(i,j)$-th element of $\Q = \bbH_{\mathcal{M}}^T \bbH_{\mathcal{M}}$, \new{and measures the orthogonality among the columns of $\bbH_{\mathcal{M}}$ using the strength of the off-diagonal terms of $\Q$, i.e., the inner product between different colums of $\bbH_{\mathcal{M}}$. Concretely, this metric is a slight variation of the $t$-averaged mutual coherence \cite{Elad2007}, with $t = 0$, where we use a slightly different way of measuring the strength of the off-diagonal terms, which is motivated by the frame potential \cite{Ranieri2014}.}

Based on the measure in \eqref{eq:measure_orthogonality}, the proposed selection scheme is given by
\begin{equation}
  \label{eq:smart_sampling}
  \mathcal{M}^{\ast} = \mathop{\text{argmin}}_{\mathcal{M} \in \mathscr{M}} \rho_{\mathcal{M}},
\end{equation}
where $\mathscr{M}$ denotes all possible choices for the set $\mathcal{M}$, with $|\mathscr{M}| = \binom{N}{M}$.
Clearly, the solution to the optimization problem \eqref{eq:smart_sampling} requires an exhaustive search over the set $\mathscr{M}$. Since $|\mathscr{M}|$ grows quickly with $N$, this brute-force solution can become unfeasible even for small network sizes. We \new{thus solve the optimization problem in \eqref{eq:smart_sampling} using a greedy approach, similarly to \cite{Ranieri2014}}. That is, given a set $\mathcal{M}^{\ast}_{n-1}$, which is composed by $n - 1$ nodes, we add new node, $m^{\ast}$, such that
\begin{equation}
  \label{eq:smart_sampling_greedy}
  m^{\ast} = \mathop{\text{argmin}}_{m \in \mathcal{N} \setminus \mathcal{M}^{\ast}_{n-1}}  \rho_{\mathcal{M}_{n}},
\end{equation}
where $\mathcal{M}_{n} = \mathcal{M}^{\ast}_{n-1} \cup m$. This yields $\mathcal{M}^{\ast}_{n} = \mathcal{M}^{\ast}_{n-1} \cup m^{\ast}$ and we keep adding nodes until $n = M$. Regarding the initial set, $\mathcal{M}^{\ast}_{1}$ could be picked randomly. A better alternative would be to pick the first two nodes, $\mathcal{M}^{\ast}_{2}$, using an exhaustive search, which is not very computationally demanding since the number of combinations, $\binom{N}{2}$, is manageable for many values of $N$. 

Three final comments are in order. First, the measure $\rho_{\mathcal{M}}$ would become unbounded if all the elements of a column of $\bbH_{\mathcal{M}}$ were zero, which could happen if the graph is very sparse and the filter is of extremely low order. In such cases, we remove the corresponding column and row of $\Q$, and compute $\rho_{\mathcal{M}}$ by replacing $\Q$ with 
 the reduced-size matrix $\bar{\Q} \in \mathbb{R}^{\bar{M} \times \bar{M}}$. The second comment is that alternative measures of orthogonality could be used, such as $\xi_{\mathcal{M}} = \max_{i \neq j} q_{ij}^2 / q_{ii} q_{jj}$, \new{which is the square of the mutual coherence \cite{Elad2007},} but our numerical experiments have shown that $\rho_{\mathcal{M}}$ performs better. Hence, in Section \ref{sec:simulations}, only $\rho_{\mathcal{M}}$ is considered. \new{The third comment is that the problem of sampling graph signals has been considered in the graph signal literature. For instance, the works in \cite{Tsitsvero2015signals,Richard_Djuric_book} considered similar metrics to that in \eqref{eq:measure_orthogonality}, but in the context of bandlimited signals.}

\subsection{Relation with the reconstruction of bandlimited signals} 

The problem presented in this section is closely related to that of recovering a bandlimited graph signal from a limited number of nodal observations, which has been extensively analyzed in the literature \cite{SamplingOrtegaICASSP14,Kovacevic2015,Marques2016Sampling,Tsitsvero2015signals,Varma2015}. As explained in Section \ref{Ss:alternative_subspace_signal_models}, bandlimited signals have a \textit{sparse frequency} representation $\|\tby\|_0=\|\bbU\bby\|_0\leq S_{\tilde{y}}$, so that $\bby=\sum_{k\in \ccalF}\bbv_k \tilde{y}_k$ with $\ccalF$ denoting the frequency support. \new{Although most works assume that the set of active frequencies is known beforehand and set to $\ccalF$, some authors have also investigated the reconstruction for a generic unknown $\ccalF$ \cite{Marques2016Sampling,Varma2015}.} For bandlimited signals, the $M$ observations in $\bby_{\ccalM}=\bbC_\ccalM\bbV\tby$ are used first to estimate the $S_{\tilde{y}}$ non-zero frequency coefficients in $\tby$. The estimated coefficients $\hat{\tby}$ are then used to recover the full signal as $\hby=\bbV\hat{\tby}$. 
The main differences between the observation models given by $\bby=\bbV\tby$ with $\|\tby\|_0\leq S_{\tilde{y}}$ and $\bby=\bbH\bbx$ with $\|\bbx\|_0\leq S_x$ are summarized next. 

First, while the subspace of bandlimited signals is spanned by a subset of the columns of $\bbV$, the one of diffused sparse signals is spanned by a subset of the columns of $\bbH$. Note that although different, both depend on the topology of the graph encoded in $\bbS$. \new{Second, except for the \textit{smooth} signals associated with a Laplacian shift \cite{Shuman2013,SamplingOrtegaICASSP14}, the underlying physical processes that generate bandlimited graph signals are not yet well understood.} Differently, diffused sparse signals have a clear physical interpretation. Third, while for bandlimited signals the estimation of $\tby$ is just an intermediate step to reconstruct the full $\bby$, in our case finding the sparse signal $\bbx$ can have practical interest too. Finally, while for diffused sparse signals  having access to some values $x_k$ of the input can be reasonable in practice, knowledge of particular non-zero frequency coefficients $\tilde{y}_k\neq0$ may be more difficult to motivate. 

\section{Recovery for known input}
\label{sec:filter_recovery}

In this section, we investigate the setup where we have access to the subsampled output $\y_{\mathcal{M}}$, the shift matrix $\S$, and the input $\x$ (e.g., we stimulate a number of regions of the brain with an external input). The goal is then to reconstruct (estimate) either the values of $\y$ at the unobserved nodes, the filter coefficients $\h$, or both. This is relevant when the dynamics governing the underlying diffusion are unknown and the goal is either to identify such dynamics or to use them to estimate the signal at non-sampled nodes. \new{Some works have looked at the problem of identifying the filter coefficients from input-output observations \cite{segarra_2017_optimal}, but setups where access to only a subset of output observations is available and the focus is on the reconstruction of the output have been mostly ignored \cite{Chen2014,Tanaka2020_spm}. }
Since the problem is once again ill-posed, we will consider two setups that render the reconstruction tractable: one where the coefficients $\h$ are sparse (Section \ref{Ss:sparse_filter_coefficients}), and another one where $\h$ belongs to a known subspace (Section \ref{Ss:filter_coefficients_belong_to_suspace}). \new{A further step along these lines could be the design of the input signal to maximize the recovery performance, but this is out of the scope of this paper.}

\subsection{Sparse filter coefficients}\label{Ss:sparse_filter_coefficients}

We start by formulating the recovery problem, which will incorporate a sparse prior on $\h$. This is useful not only in general setups, but also when the filter is known to have a finite impulse response but its order is unknown. In that case, we may use an overestimate of $L$ and impose sparsity on $\h$. 

Before proceeding with the formulation, it is convenient to rewrite \eqref{eq:filtering_frequency_KatriRhao} as $\tilde{\y} = (\Psib^T \odot \U^T )^T (\I_L \otimes \x) \h$, where $\I_L$ is the $L \times L$ identity matrix. Then, we may write the subsampled output as $\bby_{\ccalM} = \bbC_{\ccalM} \V \tilde{\y} = \bbP_{\ccalM} (\I_L \otimes \x) \h$, where $\bbP_{\ccalM} = \bbC_{\ccalM} \bbP$ and $\bbP = \V (\Psib^T \odot \U^T )^T$. This reveals that, when the input $\bbx$ is given, the operating conditions in this section are equivalent to assuming that $\bby$ lives in a subspace spanned by a subset of the columns of the $N \times L$ matrix $\bbP (\I_L \otimes \x)$. With these notational conventions, the filter estimation and the signal interpolation problem may be formulated as
\begin{align}
\label{eq:feasibility_l0_h}
\hbh= \mathop{\text{find}} \,\, \{\bbh\},  \quad \text{s. to} \,\,\, \y_{\ccalM} = \bbP_{\ccalM} (\I_L \otimes \x) \h, \,\,\, \|\h\|_{0} \leq S_h,
\end{align}
from which one can then obtain $\hby_{\ccalM^c} = \bbC_{\ccalM^c}\sum_{l=0}^{L-1} \hat{h}_l \bbS^l \bbx.$
As in the previous section, we consider the minimization form of \eqref{eq:feasibility_l0_h}, i.e.,
\begin{align}
\label{eq:minimization_l0_h}
\hbh= \mathop{\text{argmin}}_{\h} \,\,  \|\h\|_{\w,0}, \quad
\text{s. to} \,\,\, \y_{\ccalM} = \bbP_{\ccalM} (\I_L \otimes \x) \h. 
\end{align}
Note that we have exchanged the $\ell_0$ norm by the weighted $\ell_0$ norm, given by $\|\h\|_{\w,0} = \sum_{l = 0}^{L-1} w_l u(|h_l|),$
with $u(x) = 0$ if $x \leq 0$ and $u(x) = 1$ if $x > 0$. These weights allow us to promote certain coefficients. For instance, in the aforementioned scenario where the filter order is unknown, it would make sense to penalize high complexity in $\bbH$ by favoring non-zero coefficients for small $l$, that is, $w_l$ should be an increasing function in $l$. 

Since the minimization problem in \eqref{eq:minimization_l0_h} is not convex, we must use surrogates for the weighted $\ell_0$ norm, yielding $\hbh^{(i)} = \mathop{\text{argmin}}_{\h} \,\, \sum_{l = 0}^{L-1} w_l a_l |h_l|,$ subject to the same constraint, and
where $ a_l = 1,$ for the $\ell_1$ surrogate, and $a_l = (|\hat{h}_l^{(i-1)}| + \epsilon_0)^{-1}$ for the log surrogate. In the former case, only $I = 1$ iteration is necessary, whereas $I > 1$ iterations are required for the latter. 

\subsection{Filter coefficients lying on a subspace}
\label{Ss:filter_coefficients_belong_to_suspace}

Now we consider the case that $\h$ lies in a subspace $  \h = \bbD_h \bbalpha_h$, where $\bbD_h \in \mathbb{R}^{L \times D_h}$ is known and $\bbalpha_h  \in \mathbb{R}^{D_h}$. \new{One can think of the columns of $\D_h$ as containing pre-specified filters that encode our prior knowledge of potential network processes that can explain the data that we observe.}
Let us first rewrite \eqref{eq:filtering_frequency_KatriRhao} as $\tilde{\y} = (\Psib^T \odot \U^T )^T (\bbD_h \otimes \x) \bbalpha_h$, which yields the recovery problem
\begin{align}
\label{eq:feasibility_l0_h_subspace}
\hat{\bbalpha}_h= \mathop{\text{find}} \,\, \{\bbalpha_h\},  \quad
\text{s. to} \,\,\, \y_{\ccalM} = \bbP_{\ccalM} (\bbD_h \otimes \x) \bbalpha_h.
\end{align}
If the subspace is known (i.e., $\bbD_h$ is a tall matrix where all its columns are active), the feasibility problem \eqref{eq:feasibility_l0_h_subspace} is a simple system of equations that has a unique solution if $\mathrm{rank}(\bbP_{\ccalM} (\bbD_h \otimes \x)) = D_h$. 
\new{Notice that even for the case where $\D_h = \I_L$, the linear system becomes invertible for a sufficiently large number of observations.}
If the subspace is unknown, $\bbalpha_h$ can be modeled as a sparse vector. As in previous sections, the feasibility problem in \eqref{eq:feasibility_l0_h_subspace} needs to be replaced with a minimization whose objective is a (reweighted) $\ell_1$-norm promoting sparsity on $\hat{\bbalpha}_h$.

\section{Recovery for unknown filter and input}
\label{S:jointblind}

In this section, we drop some of the previous assumptions and consider that only the shift matrix $\S$ and (an upper bound on) the filter order $L-1$ are known. The goal is then to use the set of observations $\y_{\ccalM}$ to recover either the values of $\bby$ at the unobserved nodes, the input signal $\x$, or the filter coefficients $\h$. The problem in this case is more challenging since, on top of being ill-posed, the relation between the unknowns $(\bby, \bbh, \bbx)$ is bilinear. The working assumption in this section is that both $\x$ and $\h$ are sparse. We first formulate the recovery as a sparse and bilinear optimization problem and present pertinent convex relaxations (Section \ref{sec:CO_based_solution}). We then discuss the case where knowledge of $\bbx_{\ccalK}$, the values input $\bbx$ at a subset of nodes, is available. 

The recovery problem considered in this section is formulated as
\begin{align}
\label{eq:feasibility}
   \{\hbx,\hbh\}=& \, \text{find} \,\,\, \{\x,\h\}, &
  \text{s. to} \,\,\, & \y_{\ccalM} = \bbP_{\mathcal{M}} \text{vec}(\x \h^T), \,\,\, \|\x\|_0 \leq S_x, \,\,\, \|\h\|_{0} \leq S_h, 
\end{align}
from where we can obtain $\hby_{\ccalM^c} = \bbC_{\ccalM^c}\sum_{l=0}^{L-1} \hat{h}_l \bbS^l \hbx.$
The first difference between \eqref{eq:minimization_l0} or \eqref{eq:feasibility_l0_h} and \eqref{eq:feasibility} is that the number of unknowns is larger in the latter case. Concretely, there are $S_x + S_h$ versus $S_x$ or $S_h$. The second difference is that the constraints are \textit{bilinear} and not linear. The bilinear constraint is not convex and also introduces an inherent scaling ambiguity. 

To achieve a tractable relaxation, it is possible to \textit{lift} the problem by defining the $N \times L$ rank-one matrix $\bbZ = \bbx \bbh^T$, so that the problem in \eqref{eq:feasibility} becomes 
\begin{align}
\label{eq:feasibility2}
   \hbZ =\text{find} \quad &\{\Z\}, &
  \text{s. to} \quad \begin{cases} \y_{\ccalM} = \bbP_{\mathcal{M}} \text{vec}(\Z), \,\,\,\text{rank}(\Z) = 1, \\
   \|\Z\|_{2,0} \leq S_x, \,\,\, \|\Z^T\|_{2,0} \leq S_h, \end{cases}
 \end{align}
 and from the solution we can recover $\hby_{\ccalM^c} = \bbC_{\ccalM^c} \bbP \text{vec}(\Z).$
In \eqref{eq:feasibility2},  $\|\Z\|_{2,0}$ and $\|\Z^T\|_{2,0}$, defined as the number of non-zero rows and columns of $\Z$, are equivalent to $\|\x\|_0$ and $\|\h\|_0$ in \eqref{eq:feasibility}. Once the solution to the lifted problem is found, the estimates of $\x$ and $\h$ are obtained via the best rank-one approximation of $\Z$, i.e., $\hbx$ and $\hbh$ are given by the scaled left and right principal singular vectors of $\hbZ$. 

Since the problems \eqref{eq:feasibility} and \eqref{eq:feasibility2} are equivalent, so far there is no apparent benefit associated with the lifting approach. However, as we will show next, the problem in \eqref{eq:feasibility2} yields natural relaxations of the optimization problem, whereas \eqref{eq:feasibility} does not. Similar (and simpler) problems to \eqref{eq:feasibility} and \eqref{eq:feasibility2} have been recently studied in~\cite{Segarra2016,RamirezDiffusedSparse_icassp17,zhu_2020_estimating}.
 
\subsection{Algorithmic approach}
\label{sec:CO_based_solution}

We present suitable relaxations of \eqref{eq:feasibility2} that are easy to solve. The first step is to rewrite \eqref{eq:feasibility2} as a rank-minimization problem. Moreover, since $S_x$ and $S_h$ are typically unknown, we also include the corresponding terms as regularizers in the cost function. Thus, the feasibility problem becomes
\begin{align}
\label{eq:minimization_rank}
   \hbZ=\mathop{\text{argmin}}_{\Z} \quad & \text{rank}(\Z) + \tau_x \|\Z\|_{2,0} + \tau_h \|\Z^T\|_{2,0},  \\
  \text{s. to} \quad & \y_{\ccalM} = \bbP_{\mathcal{M}} \text{vec}(\Z). \nonumber
\end{align}
As we did in Section \ref{sec:filter_recovery}, it is possible to use a weighted $\ell_0$ norm to enforce certain sparsity patterns in $\h$ or $\x$. 
Guided by the idea of penalizing high order filters, we employ a weighted $\ell_0$ norm for $\bbh$. Nevertheless, it is possible to extend the formulation in a straightforward manner to include the weighted $\ell_0$ norm for $\x$. Taking this into account, and that the rank is the number of non-zero singular values, we propose to use the following cost function instead of the one in \eqref{eq:minimization_rank},
 \begin{align}
  J(\Z) 
  &= \sum_{n = 1}^{\min(N,L)} \!\!\!u\left(\sigma_n\right)  + \tau_x \!\sum_{n = 1}^{N} u\left(\|\z_n^T\|_2\right ) + \tau_h \!\sum_{l = 1}^{L} w_l u\left(\|\z_l\|_2\right),
 \end{align}
 where $\sigma_n$ is the $n$th singular value of $\Z$, and, with some abuse of notation, $\z_n^T$ and $\z_l$ denote the $n$th row and $l$th column of $\Z$, respectively. The cost function $J(\Z)$ is not convex due to the unit-step function $u(\cdot)$. Replacing $u(\cdot)$ with the absolute value yields the common nuclear norm \cite{Fazel2001} and (weighted) $\ell_{2,1}$ norm \cite{Tropp2006} convex surrogates of the rank and (weighted) $\ell_{2,0}$ norm, respectively. This approach was followed in \cite{Segarra2016} for similar problems. Here, as done in \cite{RamirezDiffusedSparse_icassp17}, we rely on the logarithm surrogate \cite{Fazel2003}, which commonly yields a better approximation, albeit it being concave and, therefore, still resulting in a non-convex problem. Concretely, we propose to approximate $J(\Z)$ as
  \begin{multline}
  \label{eq:approx_J}
  J(\Z) \approx \sum_{n = 1}^{\min(N,L)} \log\left(\sigma_n + \epsilon_1\right)  + \tau_x \sum_{n = 1}^{N} \log\left(\|\z_n^T\|_2 + \epsilon_2\right ) \\ + \tau_h \sum_{l = 1}^{L} w_l \log\left(\|\z_l\|_2 + \epsilon_3\right),
 \end{multline}
where  $\epsilon_1$, $\epsilon_2$, and $\epsilon_3$ are small positive constants. One drawback of the approximation in \eqref{eq:approx_J} is the presence of singular values of $\bbZ$, which are non-explicit functions of the optimization variables, i.e., the entries of $\bbZ$. 
To overcome this, recalling that $\bbZ$ is not a square matrix, following \cite{Fazel2003} we resort to the semidefinite embedding lemma. Hence, minimizing \eqref{eq:approx_J} subject to the linear constraint in~\eqref{eq:minimization_rank} is equivalent to
\begin{align}
\label{eq:minimization_log}  
 \mathop{\text{min}}_{\Z, \Thetab_1, \Thetab_2} \quad & \sum_{j=1}^{2} \log \det \left(\Thetab_j + \epsilon_1 \I \right) + \tau_x \sum_{n = 1}^{N} \log\left(\|\z_n^T\|_2 + \epsilon_2\right ) \nonumber \\ & + \tau_h \sum_{l = 1}^{L} w_l \log\left(\|\z_l\|_2 + \epsilon_3\right),  \\
  \text{s. to} \quad & \y_{\ccalM} = \bbP_{\mathcal{M}} \text{vec}(\Z), \,\,\,\,\,\,  \begin{bmatrix} \Thetab_1 & \Z \\ \Z^T & \Thetab_2 \end{bmatrix} \succeq \0. \nonumber
 \end{align}
As already pointed out, the optimization problem in \eqref{eq:minimization_log} is not convex due to the concavity of the logarithm. As in previous sections, the MM technique is used to circumvent this. The work in \cite{Fazel2003} proposes the first order Taylor polynomial of the logarithm as majorizing function, which yields the following semidefinite programming problem at the $i$th iteration
\begin{align}
  \label{eq:minimization_mm}
\!\!\!\!  \{\hat{\Z}^{(i)},   \hat{\Thetab}_1^{(i)}, \hat{\Thetab}_2^{(i)}\}& =  \mathop{\text{argmin}}_{\Z, \Thetab_1, \Thetab_2}  \, \sum_{j=1}^{2} \text{Tr} \left(\Deltab_j \Thetab_j\right) + \tau_x \sum_{n = 1}^{N} a_n \|\z_n^{T}\|_2 + \tau_h \sum_{l = 1}^{L} w_l b_l \|\z_l\|_2,  \\
  \text{s. to} \quad & \y_{\ccalM} = \bbP_{\mathcal{M}} \text{vec}\left(\Z\right), \,\,\,\,\,\, \begin{bmatrix} \Thetab_1 & \Z \\ \Z^T & \Thetab_2 \end{bmatrix} \succeq \0. \nonumber
\end{align}
where $\bbDelta_j = \big(\hat{\Thetab}_j^{(i-1)} + \epsilon_1 \I \big)^{-1}$ is a weight matrix, and $ a_n = (\|\hat{\z}_n^{T(i-1)}\|_2 + \epsilon_2)^{-1} $ and $ b_l = (\|\hat{\z}_l^{(i-1)}\|_2 + \epsilon_3)^{-1}$ are scalar weights. Then, the problem in \eqref{eq:minimization_mm} can be handled by any of the many available off-the-shelf solvers. If computational efficiency is a concern, ad-hoc algorithms may be devised. 
Moreover, it is important to point out that the last two terms in \eqref{eq:minimization_mm} may be seen as a reweighted $\ell_{2,1}$ norm.
 
 Let us end with a few comments. First of all, if $\Deltab_j = \I$, $a_n = b_l = 1$, the optimization problem in \eqref{eq:minimization_mm} corresponds to the nuclear norm and $\ell_{2,1}$ norm minimization problems. Secondly, \eqref{eq:minimization_mm} can be adapted to account for noise in the observations and model mismatches. In this case, it would suffice to replace the linear constraint by the convex constraint $\left\|\y_{\ccalM}  - \bbP_{\mathcal{M}} \text{vec}(\bbZ) \right\|_2^2 \leq \varepsilon$. Finally, we shall briefly comment on the recoverability of \eqref{eq:minimization_mm}. It is expected that the performance depends on $\bbP_{\mathcal{M}}$, and it is therefore driven by the sampling set $\ccalM$ and the graph topology encoded in the graph shift. Accordingly, there exist graph topologies and sampling sets that provide better performances as demonstrated in the simulations shown in Section \ref{sec:simulations}. Regarding the sampling sets, we can use as proxy for the performance the orthogonality among the columns in $\bbP_{\mathcal{M}}$ and use the criterion in \eqref{eq:smart_sampling} (or its greedy version in~\eqref{eq:smart_sampling_greedy}), where $\rho_{\mathcal{M}}$ is still given by \eqref{eq:measure_orthogonality}, with $\Q = \bbP_{\mathcal{M}}^T \bbP_{\mathcal{M}}$.

\subsection{Known input values}
\label{Ss:jointblind_knowninputs}

In this section, we discuss how to exploit the knowledge of input values at a subset of nodes. That is, we show how \eqref{eq:minimization_mm} must be modified to incorporate such information and how to modify the rank-one approximation of $\hbZ=\hbZ^{(I)}$.

Without loss of generality, we consider that the known values of $\bbx$ are non-zero. 
If they were zero, it would suffice to set to zero the corresponding rows of $\Z$ and remove them from the optimization. 
Then, taking into account that $\Z = \x \h^T$, the rows of $\Z$ are proportional to each other, that is, $\z_{i}^T/x_{i} = \h^T$ for all $i$. 
Thus, to exploit the information provided by $\x_{\ccalK}$, which collects the $K$ known (non-zero) values of $\x$, we need to augment \eqref{eq:minimization_mm} by incorporating the additional constraints
\begin{equation}\label{E:constraint_known_x}
  \z_{k_i}^{T} x_{k_{i+1}} =   \z_{k_{i+1}}^{T} x_{k_i}, \quad i = 1, \ldots, K-1,
\end{equation}
with $\ccalK = \{k_1, \ldots, k_K\}$. Since these constraints are linear, the optimization problem in \eqref{eq:minimization_mm} is still convex. This approach is only possible when there are $K \geq 2$ known values.

As we just pointed out, the information provided by $\x_{\ccalK}$ must also be exploited in the rank-one approximation of $\hbZ$, as shown in the next proposition.
\begin{proposition}\label{Prop:known_x}
  The best rank-one approximation of $\hat{\Z}$ given $\x_{\ccalK}$ is
\begin{equation}
  \begin{bmatrix} \hat{\Z}_{\mathcal{K}} \\ \hat{\Z}_{\mathcal{K}^c}\end{bmatrix}  \approx \begin{bmatrix} \x_{\mathcal{K}} \\ \hat{\x}_{\mathcal{K}^c}\end{bmatrix} \hat{\h}^T
\end{equation}
where $\hat{\x}_{\mathcal{K}^c} =\hat{\Z}_{\mathcal{K}^c} \hat{\h}/\| \hat{\h}\|^{2}$ and $ \hat{\h} =  (\check{\h}^T \hat{\Z}_{\mathcal{K}}^T \x_{\mathcal{K}}/\|\x_{\mathcal{K}}\|^2)  \check{\h},$
with $\check{\h}$ being the normalized (unit-norm) principal eigenvector of
\begin{equation}
  \label{eq:matrix_eigenvector}
  \hat{\Z}_{\mathcal{K}^c}^T  \hat{\Z}_{\mathcal{K}^c}  + \frac{1}{\|\x_{\mathcal{K}}\|^2} \hat{\Z}_{\mathcal{K}}^T \x_{\mathcal{K}} \x_{\mathcal{K}}^T \hat{\Z}_{\mathcal{K}}.
\end{equation}
\end{proposition}
\begin{proof}
The best rank-one approximation of $\hat{\Z}$ given $\x_{\ccalK}$ is obtained by minimizing $\left\|\hat{\Z} - \x \h^T\right\|_{F}^2$, subject to $ \x_{\mathcal{K}} = \C_{\mathcal{K}} \x$,
or equivalently, 
\begin{equation}
    \label{eq:minimization_rank_one_sparse}
 \mathop{\text{minimize}}_{\x_{\mathcal{K}^c},\h} \quad \left\| \begin{bmatrix} \hat{\Z}_{\mathcal{K}} \\ \hat{\Z}_{\mathcal{K}^c}\end{bmatrix}  - \begin{bmatrix} \x_{\mathcal{K}} \h^T \\ \x_{\mathcal{K}^c} \h^T\end{bmatrix} \right\|^2_F,
\end{equation}
where $\hat{\Z}_{\mathcal{K}} = \C_{\mathcal{K}} \hat{\Z}$, $\hat{\Z}_{\mathcal{K}^c} = \C_{\mathcal{K}^c} \hat{\Z}$, and $\x_{\mathcal{K}^c} = \C_{\mathcal{K}^c} \x$. Expanding the cost function in \eqref{eq:minimization_rank_one_sparse} and taking the derivative with respect to $\x_{\mathcal{K}^c}$, we obtain $\hat{\x}_{\mathcal{K}^c} = \hat{\Z}_{\mathcal{K}^c} \h / \|\h\|^2$.
Now, plugging back this solution into the cost function, \eqref{eq:minimization_rank_one_sparse} simplifies to
\begin{equation}
 \mathop{\text{maximize}}_{\h} \quad \frac{1}{\|\h\|^2} \h^T \hat{\Z}_{\mathcal{K}^c}^T  \hat{\Z}_{\mathcal{K}^c} \h + 2 \h^T \hat{\Z}_{\mathcal{K}}^T \x_{\mathcal{K}} - \|\x_{\mathcal{K}}\|^2 \|\h\|^2,
\end{equation}
and letting $\h = c_h \check{\h}$, with $\|\check{\h}\|^2 = 1$, the maximization problem becomes
\begin{align}
  \mathop{\text{maximize}}_{c_h, \check{\h}} \quad & \check{\h}^T \hat{\Z}_{\mathcal{K}^c}^T  \hat{\Z}_{\mathcal{K}^c} \check{\h} + 2 c_h \check{\h}^T \hat{\Z}_{\mathcal{K}}^T \x_{\mathcal{K}} - \|\x_{\mathcal{K}}\|^2 c_h^2, &
  \mathop{\text{s. to}} \quad & \|\check{\h}\|^2 = 1. 
\end{align}
The optimum $c_h$ is given by $\hat{c}_h = \frac{1}{\|\x_{\mathcal{K}}\|^2} \check{\h}^T \hat{\Z}_{\mathcal{K}}^T \x_{\mathcal{K}}$,
which yields
\begin{align}
  \label{eq:minimization_rank_one_no_x_unit_norm}
 \mathop{\text{maximize}}_{\check{\h}} \quad & \check{\h}^T \left( \hat{\Z}_{\mathcal{K}^c}^T  \hat{\Z}_{\mathcal{K}^c}  + \frac{1}{\|\x_{\mathcal{K}}\|^2} \hat{\Z}_{\mathcal{K}}^T \x_{\mathcal{K}} \x_{\mathcal{K}}^T \hat{\Z}_{\mathcal{K}} \right) \check{\h}, &
 \mathop{\text{s. to}} \quad & \|\check{\h}\|^2 = 1. 
\end{align}
Finally, the solution to \eqref{eq:minimization_rank_one_no_x_unit_norm} is given by the principal eigenvector of \eqref{eq:matrix_eigenvector}, which is normalized to have unit norm.
\end{proof}

In general, even if~\eqref{eq:minimization_mm} is augmented with the constraint in~\eqref{E:constraint_known_x}, the best rank-one decomposition of the obtained $\hat{\bbZ}$ will not automatically enforce the known values of $\bbx_{\ccalK}$. 
In this context, Proposition~\ref{Prop:known_x} provides a closed-form expression for a rank-one decomposition that approximates $\hat{\bbZ}$ in the best way possible while complying with the known values of $\bbx_{\ccalK}$.
Expression~\eqref{eq:matrix_eigenvector} reveals exactly how the existence of a non-empty set $\ccalK$ modifies the matrix that is used in the principal eigenvector decomposition.
Finally, notice that the knowledge of $\bbx_{\ccalK}$ also resolves the inherent problem of scale ambiguity that arises when no values are known.

\subsection{Departing from sparsity: Subspace priors}
\label{S:jointblind_subspace}

As in the previous subsections, here we also consider that only $\S$ is known, with the goal being, using $\y_{\ccalM}$, to recover the values of $\bby$ at the unobserved nodes, the input signal $\x$, and the filter coefficients $\h$. The difference with respect to Sections \ref{sec:CO_based_solution} and \ref{Ss:jointblind_knowninputs} is that, instead of assuming sparsity on the input and filter coefficients, here we assume that $\bbx$ belongs to a known subspace. That is, $\x = \bbD_x \bbalpha_x$, where $\bbalpha_x \in \mathbb {R}^{D_x}$ is a vector of coefficients and $\bbD_x \in \mathbb{R}^{N \times D_x}$ is a tall (known) matrix spanning, e.g., any of the wavelet of dictionary subspaces presented in Section \ref{Ss:alternative_subspace_signal_models}.\footnote{The method could be straightforwardly extended to the case where $\h$ belongs to a known subspace, or both do.} Under this subspace assumption for the input, \eqref{eq:filtering_frequency_KatriRhao} may be more conveniently rewritten as $\tilde{\y} =  (\Psib^T \odot (\U \bbD_x)^T )^T \text{vec}(\bbalpha_x \h^T)$, and our problem boils down to
\begin{align}
\label{eq:feasibility_subspace}
   \{\hat{\bbalpha}_x,\hbh\}=\text{find} \quad &\{\bbalpha_x,\h\}, &
  \text{s. to} \quad & \y_{\ccalM} = \T_{\mathcal{M}} \text{vec}(\bbalpha_x \h^T),  
\end{align}
where $\T_{\ccalM} = \bbC_{\ccalM} \T,$ with $\T = \V (\Psib^T \odot (\U \bbD_x)^T )^T$. 
The feasibility problem in \eqref{eq:feasibility_subspace} is analogous to that in \eqref{eq:feasibility} with the difference that there are no sparsity constraints in~\eqref{eq:feasibility_subspace}. 
Upon defining the rank-one matrix $\W = \bbalpha_x \h^T$, we may follow an approach similar to the one in Section~\ref{sec:CO_based_solution} to obtain a suitable relaxation of a rank minimization problem on $\W$.
Concretely, mimicking the steps in Section~\ref{sec:CO_based_solution} one can derive an approximate iterative solution based on MM, which at the $i$th iteration solves
\begin{align}
\label{eq:minimization_log_subspace}  
  \{\hat{\W}^{(i)},   \hat{\Thetab}_1^{(i)}, & \hat{\Thetab}_2^{(i)}\} = \,\, \mathop{\text{argmin}}_{\W, \Thetab_1, \Thetab_2}  \sum_{j=1}^{2} \text{Tr} \left(\Deltab_j \Thetab_j\right),  \\
  \text{s. to} \quad & \y_{\ccalM} = \T_{\mathcal{M}} \text{vec}(\W), \,\,\,  \begin{bmatrix} \Thetab_1 & \W \\ \W^T & \Thetab_2 \end{bmatrix} \succeq \0, \nonumber
 \end{align}
where $\bbDelta_j = \big(\hat{\Thetab}_j^{(i-1)} + \epsilon_1 \I \big)^{-1}$. Letting $\hat{\W} = \hat{\W}^{(I)}$ be the solution after convergence of the MM approach, $\hat{\bbalpha}_x$ and $ \hat{\h}$ are given by the principal left and right singular vectors of $\hat{\W}$.

\subsubsection{Known input values}

As we have done in previous sections, we show here how to exploit the information provided by known values of the input. First, we incorporate this information to \eqref{eq:minimization_log_subspace}. To address this task, consider $\W = \bbalpha_x \h^T$ and then left-multiply by $\bbD_x$ on both sides, yielding $\bbD_x \W = \bbD_x \bbalpha_x \h^T = \x \h^T.$
From this expression, it is easy to see that the rows of the rank-one matrix $\bbD_x \W$ must be proportional to each other, i.e., $\d_i^T \W/x_i = \h^T$ for all $i$,
where $\d_i^T$ is the $i$th row of $\bbD_x$. Hence, we must augment the problem in \eqref{eq:minimization_log_subspace} with
\begin{equation}
\label{E:constraint_known_x_subspace}
\d_{k_i}^T \W x_{k_{i+1}} = \d_{k_{i+1}}^T \W x_{k_{i}}, \quad i = 1, \ldots, K-1,
\end{equation}
which only works for $K \geq 2$.

The information provided by $\x_{\mathcal{K}}$ can also be used in the rank-one approximation, as shown in the following proposition.

\begin{proposition}
  \label{Prop:known_x_subspace}
  The best rank-one approximation of $\hat{\W}$ given $\x_{\ccalK}$ is $\hat{\W} \approx \hat{\bbalpha}_x \hat{\h}^T$, where $\hat{\h} = \hat{\W}^T \hat{\bbalpha}_x / \|\hat{\bbalpha}_x\|^2$, and $\hat{\bbalpha}_x$ is the principal eigenvector of
\begin{equation}
  \label{eq:matrix_eigenvector2}
  \left(\P_{\B}^{\perp} \right)^T \hat{\W} \hat{\W}^T \P_{\B}^{\perp}.
\end{equation}
Here, $\P_{\B}^{\perp} = \I_{D_s} - \B^{T} \left(\B \B^T\right)^{-1} \B$, $\B = [\I_{K-1} \ \0] \, \P_{\x_{\mathcal{K}}}^{\perp} \bbD_{\mathcal{K}}$, $\bbD_{\mathcal{K}} = \C_{\mathcal{K}} \bbD_x$, and the projector onto the orthogonal subspace to that spanned by $\x_{\mathcal{K}}$ is $\P_{\x_{\mathcal{K}}}^{\perp} = \I_K - \x_{\mathcal{K}} \x_{\mathcal{K}}^T/\|\x_{\mathcal{K}}\|^2$.
\end{proposition}

\begin{proof}
The best rank-one approximation of $\hat{\W}$ given $\x_{\ccalK}$ is obtained as the solution to the minimization problem
\begin{align}
  \label{eq:minimization_rank_one_subspace}
  \mathop{\text{minimize}}_{\bbalpha_x,\h} \,\, \left\|\hat{\W} - \bbalpha_x \h^T\right\|_{F}^2, \quad
  \mathop{\text{s. to}} \,\,\, \x_{\mathcal{K}} = \bbD_{\mathcal{K}} \bbalpha_x.
\end{align}
Since $\h$ is not constrained, the optimal solution is found by setting the derivative equal to zero, which yields $\hat{\h} = \frac{1}{\|\bbalpha_x\|^2} \hat{\W}^T \bbalpha_x$. Plugging this solution back into \eqref{eq:minimization_rank_one_subspace}, the optimization problem becomes
\begin{equation}
  \label{eq:minimization_rank_one_no_h}
  \mathop{\text{maximize}}_{\bbalpha_x} \,\,  \frac{\bbalpha_x^T \hat{\W} \hat{\W}^T \bbalpha_x}{\|\bbalpha_x\|^2}, \quad
  \mathop{\text{s. to}}  \x_{\mathcal{K}} = \bbD_{\mathcal{K}} \bbalpha_x, 
\end{equation}
which is nothing but the maximization of a Rayleigh quotient subject to a set of $K$ linear constraints \cite{Cour2006}. 
Inspired by~\cite{Cour2006}, we rewrite the constraints in \eqref{eq:minimization_rank_one_no_h} as $\P_{\x_{\mathcal{K}}}^{\perp} \bbD_{\mathcal{K}} \bbalpha_x = \P_{\x_{\mathcal{K}}}^{\perp} \x_{\mathcal{K}} = \0$,
where we have used the definition of the projector onto the orthogonal subspace to that spanned by $\x_{\mathcal{K}}$ in the statement of the proposition. Notice, however, that these $K$ constraints are not all linearly independent. 
In fact, since $\text{rank}(\P_{\x_{\mathcal{K}}}^{\perp}) = K - 1$, there are only $K - 1$ independent constraints. Then, to get rid of this extra constraint, we should project onto a $K-1$ dimensional space, for instance with $[\I_{K-1} \, \0]$, which yields $[\I_{K-1} \ \0] \, \P_{\x_{\mathcal{K}}}^{\perp} \bbD_{\mathcal{K}} \bbalpha_x = \B \bbalpha_x = \0$.
Then, \eqref{eq:minimization_rank_one_no_h} may be rewritten as follows
\begin{equation}
  \label{eq:minimization_rank_one_no_h_homo}
  \mathop{\text{maximize}}_{\bbalpha_x} \quad  \frac{\bbalpha_x^T \hat{\W} \hat{\W}^T \bbalpha_x}{\|\bbalpha_x\|^2}, \quad
  \mathop{\text{s. to}} \,\,  \B \bbalpha_x = \0, 
\end{equation}
which shows that the solution must belong to the null subspace of $\B$, and can thus be parametrized as $\bbalpha_x = \P_{\B}^{\perp} \t$,
where $\t \in \mathbb{R}^{D_s}$ and $\P_{\B}^{\perp}$ defined as in the statement of the proposition.
This change of variables allows us to drop the constraint in \eqref{eq:minimization_rank_one_no_h_homo}, which becomes
\begin{equation}
  \label{eq:minimization_rank_one_no_h_homo_t}
  \mathop{\text{maximize}}_{\t} \  \frac{\t^T \left(\P_{\B}^{\perp} \right)^T \hat{\W} \hat{\W}^T \P_{\B}^{\perp}  \t}{\t^T \left(\P_{\B}^{\perp} \right)^T \P_{\B}^{\perp} \t }.
\end{equation}
Finally, since projection matrices are idempotent, \eqref{eq:minimization_rank_one_no_h_homo_t} may be rewritten as
\begin{multline}
  \label{eq:minimization_rank_one_no_h_homo_x2}
  \mathop{\text{maximize}}_{\t} \  \frac{\t^T \left(\P_{\B}^{\perp} \right)^T \left(\P_{\B}^{\perp} \right)^T \hat{\W} \hat{\W}^T \P_{\B}^{\perp} \P_{\B}^{\perp}  \t}{\t^T \left(\P_{\B}^{\perp} \right)^T \P_{\B}^{\perp} \t } \\ = \mathop{\text{maximize}}_{\bbalpha_x} \  \frac{\bbalpha_x^T \left(\P_{\B}^{\perp} \right)^T \hat{\W} \hat{\W}^T \P_{\B}^{\perp} \bbalpha_x}{\bbalpha_x^T \bbalpha_x },
\end{multline}
and the solution is therefore given by the principal eigenvector of \eqref{eq:matrix_eigenvector2}.
\end{proof}

Similarly to the previous subsection, augmenting \eqref{eq:minimization_log_subspace}  with \eqref{E:constraint_known_x_subspace} does not suffice to enforce the known values of $\bbx_{\ccalK}$. This is achieved via Proposition~\ref{Prop:known_x_subspace}, which provides the best rank-one approximation of $\hat{\bbW}$ that exploits the known values of $\bbx_{\ccalK}$. This proposition shows how these known values modify the matrix that is used in the principal eigenvalue decomposition.

\section{Numerical results}
\label{sec:simulations}


\begin{figure*}[t]
     \centering
     \subfloat[][]{\includegraphics{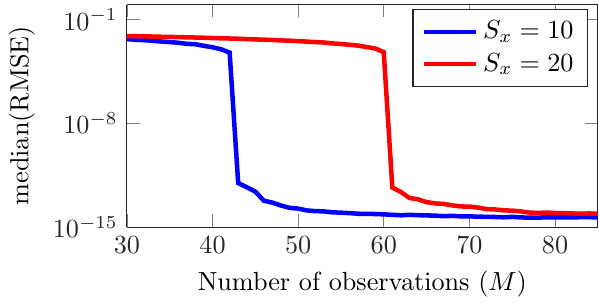}\label{fig:known_h_MSE}} \hfill
     \subfloat[][]{\includegraphics{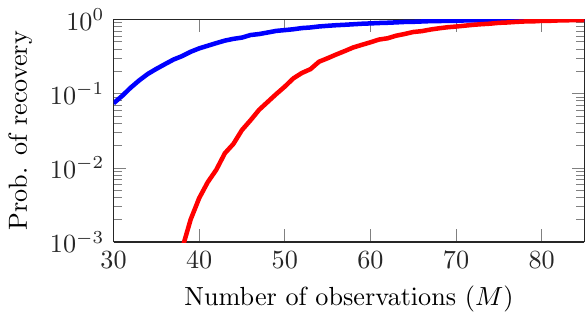}\label{fig:known_h_Pr}}
     
     \subfloat[][]{\includegraphics{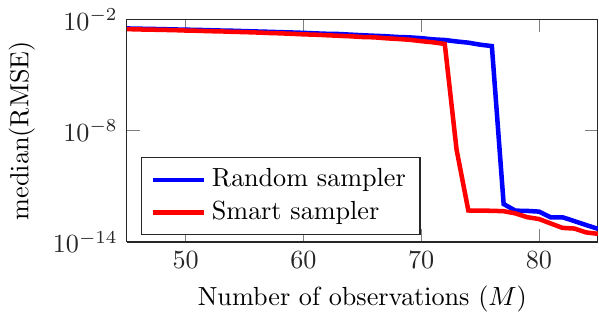}\label{fig:known_h_greedy_MSE}} \hfill
     \subfloat[][]{\includegraphics{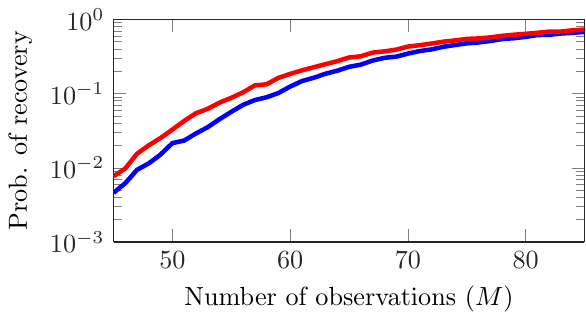}\label{fig:known_h_greedy_Pr}} 
     
     \subfloat[][]{\includegraphics{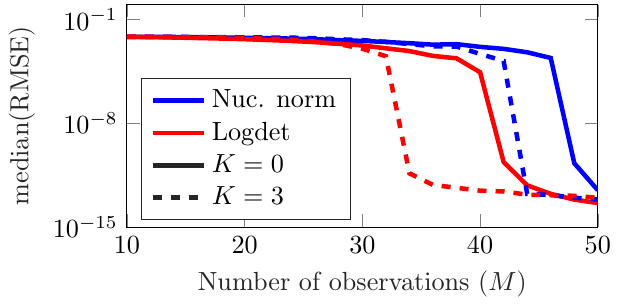}\label{fig:sparse_x_MSE}} \hfill
     \subfloat[][]{\includegraphics{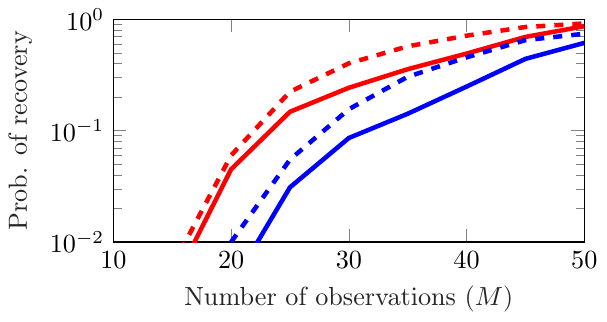}\label{fig:sparse_x_Pr}} 
     
     \caption{\small Recovery performance (left column: median of the root-mean-squared error; right column: probability of recovery) for a varying number of observations, $M$. Panels (a) and (b): Scenario with $N = 100$, known $\miH$, $L = 4$, $K = 0$, different sparsity values, and random sampling. Panels (c) and (d): Scenario with $N = 200$, known $\miH$, $L = 3$, $K = 0$, and $S_x = 15$. Panels (e) and (f): Scenario with $N = 50$, unknown $\miH$, $L = 3$, $S_x = 8$, and $S_h = L$.}
\end{figure*}

The main goal of this section is to numerically analyze the behavior of our algorithms in a range of setups, assessing the impact of different parameters in the recovery performance. Due to the plethora of possible scenarios, we present here a selection of the ones that are more interesting and insightful.
Unless otherwise stated, all scenarios consider an Erd\H{o}s-R\'enyi random graph \cite{bollobas1998random} with edge-presence probability $p = 0.1$, and the obtained adjacency matrix is the considered choice for the shift matrix. The (non-zero) values of $\x$ or $\bbalpha_x$ and $\h$ are drawn from a multivariate Gaussian distribution and normalized afterwards to have unit norm. Finally, the entries of $\bbD_x$ are independent and identically distributed Gaussian variables with zero mean and unit variance, and the whole matrix $\bbD_x$ is then normalized to have unit Frobenius norm.

\subsection{Recovery with known diffusing filter}

First, we consider the simpler scenario where the graph filter $\bbH$ is perfectly known. 
For this case, we have run $10^4$ Monte Carlo simulations and in each of them, new realizations of the graph $\ccalG$, the signal $\bbx$, and the coefficients $\bbh$ are drawn. The performance of the log surrogate technique in \eqref{eq:minimization_reweighted_l11} is measured by the median of the normalized root-mean-squared error, which is defined as $\mathrm{RMSE} = \|\hat{\x} - \x \|_2/N,$
and also by the probability of recovery, given by $\text{Pr}(\mathrm{RMSE} \leq 10^{-5}).$

Figs. \ref{fig:known_h_MSE} and \ref{fig:known_h_Pr} show these metrics for varying $M$ and random sampling in a scenario with a graph of size $N = 100$ nodes, a filter with $L = 4$ coefficients, no known input values, and $S_x \in \{10,20\}$. As expected, the value of $S_x$ markedly affects the recoverability. 

Next, we analyze the performance improvement provided by the sampling technique proposed in Section~\ref{Sec:optimizing_sampling}. Figs. \ref{fig:known_h_greedy_MSE} and \ref{fig:known_h_greedy_Pr} show the median of the RMSE and the probability of recovery for varying $M$ in an example with $N = 200$, $L = 3$, no known values of $\x$, and $S_x = 15$. 
Given the size of the graph, we have implemented the greedy version \eqref{eq:smart_sampling_greedy}. Concretely, we can see that the sampling technique achieves the same performance of the random sampling with around $5$ observations less in this particular setup.

\begin{figure*}[t]
     \centering
     \subfloat[][]{\includegraphics{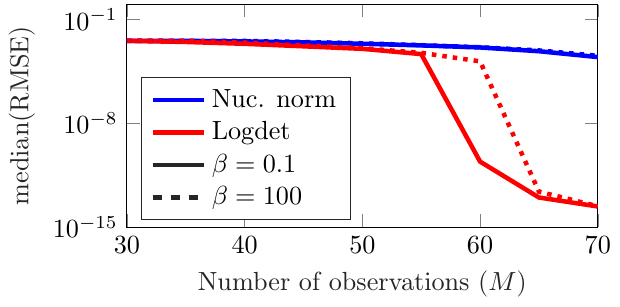}\label{fig:double_sparse_MSE}} \hfill
     \subfloat[][]{\includegraphics{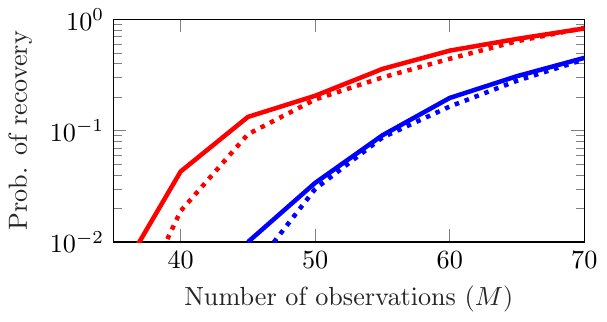}\label{fig:double_sparse_Pr}}
     
     \subfloat[][]{\includegraphics{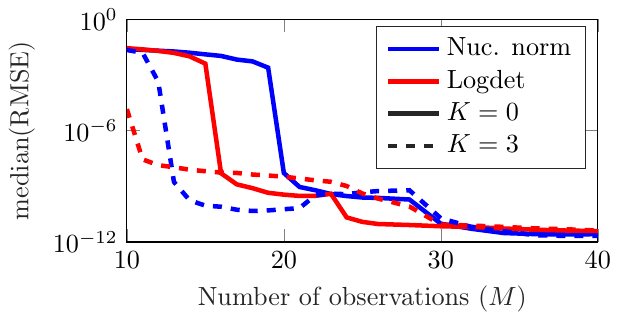}\label{fig:subspace_MSE}} \hfill
     \subfloat[][]{\includegraphics{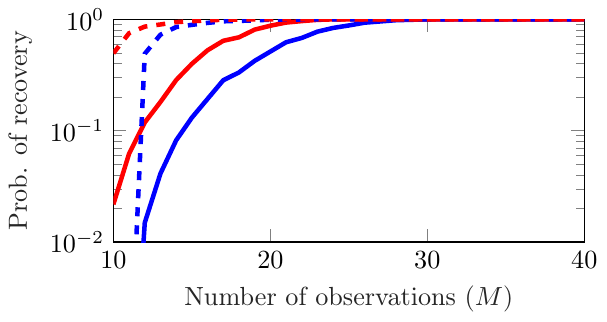}\label{fig:subspace_Pr}}
     
     \subfloat[][]{\includegraphics{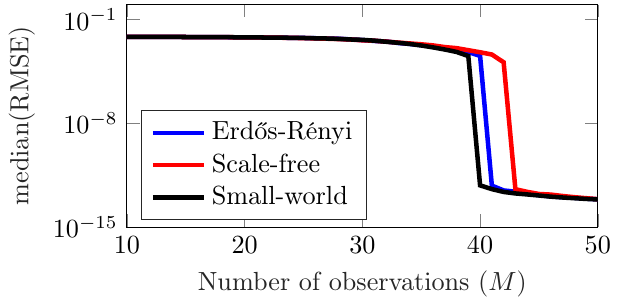}\label{fig:Different_networks_MSE}} \hfill
     \subfloat[][]{\includegraphics{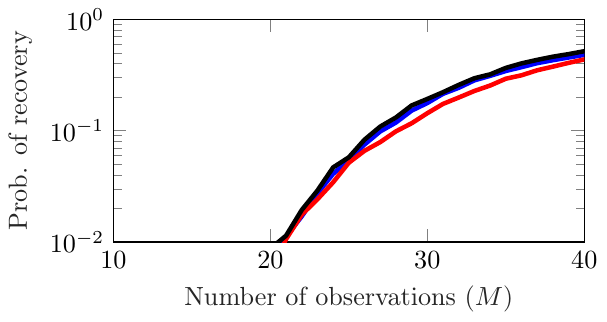}\label{fig:Different_networks_Pr}}
     
     \caption{\small Recovery performance (left column: median of the root-mean-squared error; right column: probability of recovery) for a varying number of observations, $M$. Panels (a) and (b): Scenario with $N = 70$, unknown $\miH$, $L = 4$, $S_x = 8$, $K = 0$, and $S_h = 2$. Panels (c) and (d): Scenario with $N = 50$, unknown $\miH$, $L = 3$, and $D_s = 10$. Panels (e) and (f): Scenario with $N = 50$, unknown $\miH$, $L = 3$, $S_x = 8$, and $S_h = L$.}
\end{figure*}

\subsection{Recovery with unknown diffusing filter: Sparse model}

We now evaluate the performance of the presented recovery techniques for the case of sparse input signal and filter coefficients. As in the previous section, the performance is measured by the probability of recovery, defined as above, and by the $\mathrm{RMSE} = \|\hat{\x} \hat{\h}^T - \x \h^T\|_F/(N - K)L,$
with $\| \cdot \|_F$ denoting the Frobenius norm. In this case, these measures are obtained by averaging $10^3$ Monte Carlo realizations of the graph, signal, and filter coefficients.

The results of the first experiment are shown in Figs. \ref{fig:sparse_x_MSE} and \ref{fig:sparse_x_Pr}, where we consider a scenario with $N = 50$ nodes, $S_x = 8$, and $L = S_h = 3$ filter coefficients, i.e., the filter coefficients are not sparse. Moreover, we have compared the performance of the nuclear norm and $\ell_{2,1}$ norm surrogates with that of the logdet and reweighted $\ell_{2,1}$ norm for $K = 0$ and $K = 3$ known values of $\x$. In particular, the parameters selected for both techniques are $\tau_x = 0.1$ and $\tau_h = 0$, i.e., we do not enforce sparsity in the filter coefficients. It is clear that the performance of the log surrogates is much better than that of the nuclear and $\ell_{2,1}$ norms, and can achieve the same performance with up to $10$ observations less in the best case for this scenario. Additionally, the knowledge of as few as $3$ known input values may provide a significant performance boost and the technique that exploits this information may require many less observations than the number of known values to achieve the same performance.

In the second example for the sparse model, we consider that the filter order is not known and only an overestimate is available. 
In this case, we will proceed as if the filter coefficients were a sparse vector and let the method infer that the support is concentrated at the first entries of the vector $\h$. To achieve that, we set $\tau_h = 0.1$ for both approaches and we pick exponential weights, that is, $w_l = (1 - e^{-\beta l})/(1 - e^{-\beta L}),  l = 1, \ldots, L,$
with $\beta \in \{0.1, 100\}$. The second value of $\beta$ corresponds to the case where all weights are (almost) identical. The performance of the techniques is shown in Figs. \ref{fig:double_sparse_MSE} and \ref{fig:double_sparse_Pr} for a scenario with $N = 70$ nodes, no known input values, $S_x = 8$, true filter order $S_h = 2$, which is unknown, and its overestimate given by $L =4$. Moreover, $\tau_x = 0.1$ as in the previous example. 
We can see from this figure that the exponential weights ($\beta = 0.1$) provide an advantage over the (almost) constant weights ($\beta = 100$), since the former better encodes our prior on the support of the filter coefficients.

\subsection{Recovery with unknown diffusing filters: Subspace model}

Here we evaluate the performance of the proposed recovery algorithm when the signal adheres to the subspace model.  Again, the probability of recovery and the $\mathrm{RMSE} = \|\hat{\bbalpha} \hat{\h}^T - \bbalpha \h^T\|_F/(D_s - K)L$, 
are the performance metrics, which are obtained with $10^3$ Monte Carlo simulations. Concretely, this experiment considers a graph with $N = 50$ nodes, $L = 3$, $D_s = 10$, and two values for the number of known values of $\x$: $K = 0$ and $K = 3$. Under this signal model, Figs. \ref{fig:subspace_MSE} and \ref{fig:subspace_Pr} show that the log surrogates still provide the best recovery, and the use of few known input values greatly boosts the performance.  

\subsection{Effect of the network topology}

The topology of the graph on which the diffusion occurs naturally affects the difficulty of the recovery task at hand. To illustrate this, we analyze the performance of the recovery technique in \eqref{eq:minimization_mm}, i.e., the case of sparse input signal and filter coefficients with logdet surrogate, for three different types of networks: Erd\H{o}s-R\'enyi with edge-presence probability $p = 0.12$, a scale-free (Barab\'asi-Albert) with $7$ initial nodes and where each new node is connected to $3$ existing nodes, and small-world (Watts-Strogatz) with mean node degree $6$ and rewiring probability $1$. 
For the scale-free and small-world networks, their parameters have been chosen to yield a number of links similar to those expected in the Erd\H{o}s-R\'enyi network. Figs. \ref{fig:Different_networks_MSE} and \ref{fig:Different_networks_Pr} show the recovery performance in a scenario with $N = 50$ nodes, $S_x = 8$, $S_h = L = 3$ and no known values. The algorithm parameters are $\tau_x = 0.1$ and $\tau_h = 0$, since the filter coefficients vector is not sparse. In these figures, we can see that small-world networks are easier to identify, followed by Erd\H{o}s-R\'enyi and scale-free ones.
\new{We can intuitively understand why the recovery performance on small-world graphs is the best as follows. 
	To recover the sources, the ideal scenario is to have a graph consisting of tight cliques, so that the information from the sources is maintained locally even after diffusion.
	As an extreme, if we consider the empty graph where the information is not diffused at all, it becomes trivial to determine the support of the input.
	However, in such a case, our observations on the nodes that were not sources, would be uninformative.
	Hence, it would be impossible to recover the magnitudes of the sources and the coefficients of the filter.
	On the other hand, allowing for fast diffusion across the graph implies that the observation at every node is informative, even for low-degree graph filters.
	In this way, small-world graphs combine the best of both worlds, by having high clustering coefficients (so that sources are easy to identify) and low diameter (so that more observations are informative about the filter coefficients).}

\section{Conclusions}
\label{S:Conclusions}

This paper presented a range of recovery and inverse problems for graph signals sampled at a subset of nodes. Assuming that the graph \new{shift} is known and the observed signal can be modeled as the output of a graph filter, the goal was to use the values of the signal in a few nodes to estimate the unknown input/output/filter values, encompassing and generalizing many of the schemes existing in the literature. Different setups were considered, including those of sparse signal and channel coefficients, and subspace models. In all the cases, the problems were formulated as graph-aware sparse and low-rank reconstructions, which were then relaxed to give rise to tractable approximated algorithms.  Applications of interest include graph signal interpolation and denoising, network source localization, and estimating the diffusion dynamics of a network process. Ongoing work includes the generalization of the proposed schemes to setups where the graph itself is not (or only partially) known.


\section*{Acknowledgements}

The work of the last two authors was supported by the Spanish MINECO grant PID2019-105032GB-I00 (SPGraph). The work of the first author was supported by the Ministerio de Ciencia, Innovaci{\'o}n y Universidades under grant TEC2017-92552-EXP (aMBITION), by the Ministerio de Ciencia, Innovaci{\'o}n y Universidades, jointly with the European Commission (ERDF), under grant TEC2017-86921-C2-2-R (CAIMAN), and by The Comunidad de Madrid under grant Y2018/TCS-4705 (PRACTICO-CM). 



\bibliographystyle{elsarticle-num-names}
\bibliography{arxiv}

\begin{thebibliography}{40}
\expandafter\ifx\csname natexlab\endcsname\relax\def\natexlab#1{#1}\fi
\providecommand{\url}[1]{\texttt{#1}}
\providecommand{\href}[2]{#2}
\providecommand{\path}[1]{#1}
\providecommand{\DOIprefix}{doi:}
\providecommand{\ArXivprefix}{arXiv:}
\providecommand{\URLprefix}{URL: }
\providecommand{\Pubmedprefix}{pmid:}
\providecommand{\doi}[1]{\href{http://dx.doi.org/#1}{\path{#1}}}
\providecommand{\Pubmed}[1]{\href{pmid:#1}{\path{#1}}}
\providecommand{\bibinfo}[2]{#2}
\ifx\xfnm\relax \def\xfnm[#1]{\unskip,\space#1}\fi
\bibitem[{Shuman et~al.(2013)Shuman, Narang, Frossard, Ortega, and
  Vandergheynst}]{Shuman2013}
\bibinfo{author}{D.~I. Shuman}, \bibinfo{author}{S.~K. Narang},
  \bibinfo{author}{P.~Frossard}, \bibinfo{author}{A.~Ortega},
  \bibinfo{author}{P.~Vandergheynst},
\newblock \bibinfo{title}{The emerging field of signal processing on graphs:
  Extending high-dimensional data analysis to networks and other irregular
  domains},
\newblock \bibinfo{journal}{IEEE Signal Process. Mag.} \bibinfo{volume}{30}
  (\bibinfo{year}{2013}) \bibinfo{pages}{83--98}.
\bibitem[{Sandryhaila and Moura(2014)}]{Sandryhaila2014}
\bibinfo{author}{A.~Sandryhaila}, \bibinfo{author}{J.~M.~F. Moura},
\newblock \bibinfo{title}{Big data analysis with signal processing on graphs:
  Representation and processing of massive data sets with irregular structure},
\newblock \bibinfo{journal}{IEEE Signal Process. Mag.} \bibinfo{volume}{31}
  (\bibinfo{year}{2014}) \bibinfo{pages}{80--90}.
\bibitem[{Ortega et~al.(2018)Ortega, Frossard, Kovacevi{\'c}, Moura, and
  Vandergheynst}]{GSP_overview2017}
\bibinfo{author}{A.~Ortega}, \bibinfo{author}{P.~Frossard},
  \bibinfo{author}{J.~Kovacevi{\'c}}, \bibinfo{author}{J.~M.~F. Moura},
  \bibinfo{author}{P.~Vandergheynst},
\newblock \bibinfo{title}{Graph signal processing: Overview, challenges, and
  applications},
\newblock \bibinfo{journal}{Proc. IEEE} \bibinfo{volume}{106}
  (\bibinfo{year}{2018}) \bibinfo{pages}{808--828}.
\bibitem[{Chen et~al.(2014)Chen, Sandryhaila, Moura, and Kovacevic}]{Chen2014}
\bibinfo{author}{S.~Chen}, \bibinfo{author}{A.~Sandryhaila},
  \bibinfo{author}{J.~M.~F. Moura}, \bibinfo{author}{J.~Kovacevic},
\newblock \bibinfo{title}{{Signal recovery on graphs: Variation minimization}},
\newblock \bibinfo{journal}{IEEE Trans. Signal Process.} \bibinfo{volume}{63}
  (\bibinfo{year}{2014}) \bibinfo{pages}{4609--4624}.
\bibitem[{Chamon and Ribeiro(2018)}]{Chamon2018}
\bibinfo{author}{L.~F.~O. Chamon}, \bibinfo{author}{A.~Ribeiro},
\newblock \bibinfo{title}{Greedy sampling of graph signals},
\newblock \bibinfo{journal}{IEEE Trans. Signal Process.} \bibinfo{volume}{66}
  (\bibinfo{year}{2018}) \bibinfo{pages}{34--47}.
\bibitem[{Sakiyama et~al.(2019)Sakiyama, Tanaka, Tanaka, and
  Ortega}]{Sakiyama2019}
\bibinfo{author}{A.~Sakiyama}, \bibinfo{author}{Y.~Tanaka},
  \bibinfo{author}{T.~Tanaka}, \bibinfo{author}{A.~Ortega},
\newblock \bibinfo{title}{Eigendecomposition-free sampling set selection for
  graph signals},
\newblock \bibinfo{journal}{IEEE Trans. Signal Process.} \bibinfo{volume}{67}
  (\bibinfo{year}{2019}) \bibinfo{pages}{2679--2692}.
\bibitem[{Tanaka et~al.(2020)Tanaka, Eldar, Ortega, and
  Cheung}]{Tanaka2020_spm}
\bibinfo{author}{Y.~Tanaka}, \bibinfo{author}{Y.~C. Eldar},
  \bibinfo{author}{A.~Ortega}, \bibinfo{author}{G.~Cheung},
\newblock \bibinfo{title}{Sampling signals on graphs: From theory to
  applications},
\newblock \bibinfo{journal}{IEEE Signal Process. Mag.} \bibinfo{volume}{37}
  (\bibinfo{year}{2020}) \bibinfo{pages}{14--30}.
\bibitem[{Anis et~al.(2014)Anis, Gadde, and Ortega}]{SamplingOrtegaICASSP14}
\bibinfo{author}{A.~Anis}, \bibinfo{author}{A.~Gadde},
  \bibinfo{author}{A.~Ortega},
\newblock \bibinfo{title}{Towards a sampling theorem for signals on arbitrary
  graphs},
\newblock in: \bibinfo{booktitle}{IEEE Int. Conf. on Acoustics, Speech and
  Signal Process.}, \bibinfo{year}{2014}, pp. \bibinfo{pages}{3864--3868}.
\bibitem[{Chen et~al.(2015)Chen, Varma, Sandryhaila, and
  Kovacevi{\'c}}]{Kovacevic2015}
\bibinfo{author}{S.~Chen}, \bibinfo{author}{R.~Varma},
  \bibinfo{author}{A.~Sandryhaila}, \bibinfo{author}{J.~Kovacevi{\'c}},
\newblock \bibinfo{title}{{Discrete signal processing on graphs: Sampling
  theory}},
\newblock \bibinfo{journal}{IEEE Trans. Signal Process.} \bibinfo{volume}{63}
  (\bibinfo{year}{2015}) \bibinfo{pages}{6510--6523}.
\bibitem[{Marques et~al.(2016)Marques, Segarra, Leus, and
  Ribeiro}]{Marques2016Sampling}
\bibinfo{author}{A.~G. Marques}, \bibinfo{author}{S.~Segarra},
  \bibinfo{author}{G.~Leus}, \bibinfo{author}{A.~Ribeiro},
\newblock \bibinfo{title}{Sampling of graph signals with successive local
  aggregations},
\newblock \bibinfo{journal}{IEEE Trans. Signal Process.} \bibinfo{volume}{64}
  (\bibinfo{year}{2016}) \bibinfo{pages}{1832 -- 1843}.
\bibitem[{Tsitsvero et~al.(2016)Tsitsvero, Barbarossa, and
  Lorenzo}]{Tsitsvero2015signals}
\bibinfo{author}{M.~Tsitsvero}, \bibinfo{author}{S.~Barbarossa},
  \bibinfo{author}{P.~D. Lorenzo},
\newblock \bibinfo{title}{Signals on graphs: Uncertainty principle and
  sampling},
\newblock \bibinfo{journal}{IEEE Trans. Signal Process.} \bibinfo{volume}{64}
  (\bibinfo{year}{2016}) \bibinfo{pages}{4845--4860}.
\bibitem[{Lorenzo et~al.(2018{\natexlab{a}})Lorenzo, Barbarossa, and
  Banelli}]{Richard_Djuric_book}
\bibinfo{author}{P.~D. Lorenzo}, \bibinfo{author}{S.~Barbarossa},
  \bibinfo{author}{P.~Banelli},
\newblock \bibinfo{title}{Sampling and recovery of graph signals},
\newblock in: \bibinfo{booktitle}{Cooperative and graph signal processing:
  Principles and applications}, \bibinfo{publisher}{Academic Press},
  \bibinfo{year}{2018}{\natexlab{a}}.
\bibitem[{Lorenzo et~al.(2018{\natexlab{b}})Lorenzo, Banelli, Isufi,
  Barbarossa, and Leus}]{DiLorenzo2018}
\bibinfo{author}{P.~D. Lorenzo}, \bibinfo{author}{P.~Banelli},
  \bibinfo{author}{E.~Isufi}, \bibinfo{author}{S.~Barbarossa},
  \bibinfo{author}{G.~Leus},
\newblock \bibinfo{title}{Adaptive graph signal processing: Algorithms and
  optimal sampling strategies},
\newblock \bibinfo{journal}{IEEE Trans. Signal Process.} \bibinfo{volume}{66}
  (\bibinfo{year}{2018}{\natexlab{b}}) \bibinfo{pages}{3584--3598}.
\bibitem[{Loukas et~al.(2015)Loukas, Simonetto, and Leus}]{Loukas2015}
\bibinfo{author}{A.~Loukas}, \bibinfo{author}{A.~Simonetto},
  \bibinfo{author}{G.~Leus},
\newblock \bibinfo{title}{Distributed autoregressive moving average graph
  filters},
\newblock \bibinfo{journal}{IEEE Signal Process. Lett.} \bibinfo{volume}{22}
  (\bibinfo{year}{2015}) \bibinfo{pages}{1931--1935}.
\bibitem[{Segarra et~al.(2017)Segarra, Marques, and
  Ribeiro}]{segarra_2017_optimal}
\bibinfo{author}{S.~Segarra}, \bibinfo{author}{A.~G. Marques},
  \bibinfo{author}{A.~Ribeiro},
\newblock \bibinfo{title}{Optimal graph-filter design and applications to
  distributed linear network operators},
\newblock \bibinfo{journal}{IEEE Trans. Signal Process.} \bibinfo{volume}{65}
  (\bibinfo{year}{2017}) \bibinfo{pages}{4117--4131}.
\bibitem[{Sandryhaila and Moura(2014)}]{Sandryhaila2014a}
\bibinfo{author}{A.~Sandryhaila}, \bibinfo{author}{J.~M.~F. Moura},
\newblock \bibinfo{title}{Discrete signal processing on graphs: Frequency
  analysis},
\newblock \bibinfo{journal}{IEEE Trans. Signal Process.} \bibinfo{volume}{62}
  (\bibinfo{year}{2014}) \bibinfo{pages}{3042--3054}.
\bibitem[{Marques et~al.(2017)Marques, Segarra, Leus, and
  Ribeiro}]{Marques2016}
\bibinfo{author}{A.~G. Marques}, \bibinfo{author}{S.~Segarra},
  \bibinfo{author}{G.~Leus}, \bibinfo{author}{A.~Ribeiro},
\newblock \bibinfo{title}{Stationary graph processes and spectral estimation},
\newblock \bibinfo{journal}{IEEE Trans. Signal Process.} \bibinfo{volume}{65}
  (\bibinfo{year}{2017}) \bibinfo{pages}{5911--5926}.
\bibitem[{Segarra et~al.(2017)Segarra, Mateos, Marques, and
  Ribeiro}]{Segarra2016}
\bibinfo{author}{S.~Segarra}, \bibinfo{author}{G.~Mateos},
  \bibinfo{author}{A.~G. Marques}, \bibinfo{author}{A.~Ribeiro},
\newblock \bibinfo{title}{Blind identification of graph filters},
\newblock \bibinfo{journal}{IEEE Trans. Signal Process.} \bibinfo{volume}{65}
  (\bibinfo{year}{2017}) \bibinfo{pages}{1146--1159}.
\bibitem[{Iwata et~al.(2020)Iwata, Yamada, and Tanaka}]{iwata2020graph}
\bibinfo{author}{K.~Iwata}, \bibinfo{author}{K.~Yamada},
  \bibinfo{author}{Y.~Tanaka}, \bibinfo{title}{Graph blind deconvolution with
  sparseness constraint}, \bibinfo{year}{2020}.
  \href{http://arxiv.org/abs/2010.14002}{{\tt arXiv:2010.14002}}.
\bibitem[{Baraniuk(2007)}]{Baraniuk2007}
\bibinfo{author}{R.~G. Baraniuk},
\newblock \bibinfo{title}{Compressive sensing},
\newblock \bibinfo{journal}{IEEE Signal Process. Mag.} \bibinfo{volume}{24}
  (\bibinfo{year}{2007}) \bibinfo{pages}{118--121}.
\bibitem[{Ranieri et~al.(2014)Ranieri, Chebira, and Vetterli}]{Ranieri2014}
\bibinfo{author}{J.~Ranieri}, \bibinfo{author}{A.~Chebira},
  \bibinfo{author}{M.~Vetterli},
\newblock \bibinfo{title}{Near-optimal sensor placement for linear inverse
  problems},
\newblock \bibinfo{journal}{IEEE Trans. Signal Process.} \bibinfo{volume}{62}
  (\bibinfo{year}{2014}) \bibinfo{pages}{1135--1146}.
\bibitem[{Elad(2007)}]{Elad2007}
\bibinfo{author}{M.~Elad},
\newblock \bibinfo{title}{Optimized projections for compressed sensing},
\newblock \bibinfo{journal}{IEEE Trans. Signal Process.} \bibinfo{volume}{55}
  (\bibinfo{year}{2007}) \bibinfo{pages}{5695--5702}.
\bibitem[{Duarte-Carvajalino and Sapiro(2009)}]{Learning_sense_Duarte}
\bibinfo{author}{J.~M. Duarte-Carvajalino}, \bibinfo{author}{G.~Sapiro},
\newblock \bibinfo{title}{Learning to sense sparse signals: Simultaneous
  sensing matrix and sparsifying dictionary optimization},
\newblock \bibinfo{journal}{IEEE Trans. Signal Process.} \bibinfo{volume}{18}
  (\bibinfo{year}{2009}) \bibinfo{pages}{1395--1408}.
\bibitem[{Sandryhaila and Moura(2013)}]{Sandryhaila2013}
\bibinfo{author}{A.~Sandryhaila}, \bibinfo{author}{J.~M.~F. Moura},
\newblock \bibinfo{title}{Discrete signal processing on graphs},
\newblock \bibinfo{journal}{IEEE Trans. Signal Process.} \bibinfo{volume}{61}
  (\bibinfo{year}{2013}) \bibinfo{pages}{1644--1656}.
\bibitem[{Kramer et~al.(2008)Kramer, Kolaczyk, and
  Kirsch}]{kramer2008_emergent}
\bibinfo{author}{M.~A. Kramer}, \bibinfo{author}{E.~D. Kolaczyk},
  \bibinfo{author}{H.~E. Kirsch},
\newblock \bibinfo{title}{Emergent network topology at seizure onset in
  humans},
\newblock \bibinfo{journal}{Epilepsy Research} \bibinfo{volume}{79}
  (\bibinfo{year}{2008}) \bibinfo{pages}{173--186}.
\bibitem[{{Mathur} and {Chakka}(2020)}]{mathur2020_graph}
\bibinfo{author}{P.~{Mathur}}, \bibinfo{author}{V.~K. {Chakka}},
\newblock \bibinfo{title}{Graph signal processing of {EEG} signals for
  detection of epilepsy},
\newblock in: \bibinfo{booktitle}{IEEE Int. Conf. on Signal Process. and Integ.
  Networks}, \bibinfo{year}{2020}, pp. \bibinfo{pages}{839--843}.
\bibitem[{Shuman et~al.(2016)Shuman, Ricaud, and
  Vandergheynst}]{shuman2016vertex}
\bibinfo{author}{D.~I. Shuman}, \bibinfo{author}{B.~Ricaud},
  \bibinfo{author}{P.~Vandergheynst},
\newblock \bibinfo{title}{Vertex-frequency analysis on graphs},
\newblock \bibinfo{journal}{Applied and Computational Harmonic Analysis}
  \bibinfo{volume}{40} (\bibinfo{year}{2016}) \bibinfo{pages}{260--291}.
\bibitem[{Coifman and Maggioni(2006)}]{coifman2006diffusion}
\bibinfo{author}{R.~R. Coifman}, \bibinfo{author}{M.~Maggioni},
\newblock \bibinfo{title}{Diffusion wavelets},
\newblock \bibinfo{journal}{Applied and Computational Harmonic Analysis}
  \bibinfo{volume}{21} (\bibinfo{year}{2006}) \bibinfo{pages}{53--94}.
\bibitem[{Hammond et~al.(2011)Hammond, Vandergheynst, and
  Gribonval}]{hammond2011wavelets}
\bibinfo{author}{D.~K. Hammond}, \bibinfo{author}{P.~Vandergheynst},
  \bibinfo{author}{R.~Gribonval},
\newblock \bibinfo{title}{Wavelets on graphs via spectral graph theory},
\newblock \bibinfo{journal}{Applied and Computational Harmonic Analysis}
  \bibinfo{volume}{30} (\bibinfo{year}{2011}) \bibinfo{pages}{129--150}.
\bibitem[{Yankelevsky and Elad(2016)}]{yankelevsky2016dual}
\bibinfo{author}{Y.~Yankelevsky}, \bibinfo{author}{M.~Elad},
\newblock \bibinfo{title}{Dual graph regularized dictionary learning},
\newblock \bibinfo{journal}{IEEE Trans. Signal Info. Process. Networks}
  \bibinfo{volume}{2} (\bibinfo{year}{2016}) \bibinfo{pages}{611--624}.
\bibitem[{Cand{\`{e}}s et~al.(2008)Cand{\`{e}}s, Wakin, and Boyd}]{Candes2008}
\bibinfo{author}{E.~J. Cand{\`{e}}s}, \bibinfo{author}{M.~B. Wakin},
  \bibinfo{author}{S.~P. Boyd},
\newblock \bibinfo{title}{Enhancing sparsity by reweighted $l_1$ minimization},
\newblock \bibinfo{journal}{Journal of Fourier Analysis and Applications}
  \bibinfo{volume}{14} (\bibinfo{year}{2008}) \bibinfo{pages}{877--905}.
\bibitem[{Sun et~al.(2017)Sun, Babu, and Palomar}]{Palomar_MM}
\bibinfo{author}{Y.~Sun}, \bibinfo{author}{P.~Babu}, \bibinfo{author}{D.~P.
  Palomar},
\newblock \bibinfo{title}{Majorization-minimization algorithms in signal
  processing, communications, and machine learning},
\newblock \bibinfo{journal}{IEEE Trans. Signal Process.} \bibinfo{volume}{65}
  (\bibinfo{year}{2017}) \bibinfo{pages}{794--816}.
\bibitem[{Varma et~al.(2015)Varma, Chen, and Kovacevi{\'c}}]{Varma2015}
\bibinfo{author}{R.~Varma}, \bibinfo{author}{S.~Chen},
  \bibinfo{author}{J.~Kovacevi{\'c}},
\newblock \bibinfo{title}{{Spectrum-blind signal recovery on graphs}},
\newblock in: \bibinfo{booktitle}{IEEE Intl. Wrksp. Computat. Advances
  Multi-Sensor Adaptive Process.}, \bibinfo{address}{Cancun, Mexico},
  \bibinfo{year}{2015}, pp. \bibinfo{pages}{81--84}.
\bibitem[{Ram{\'\i}rez et~al.(2017)Ram{\'\i}rez, Marques, and
  Segarra}]{RamirezDiffusedSparse_icassp17}
\bibinfo{author}{D.~Ram{\'\i}rez}, \bibinfo{author}{A.~G. Marques},
  \bibinfo{author}{S.~Segarra},
\newblock \bibinfo{title}{Graph-signal reconstruction and blind deconvolution
  for diffused sparse inputs},
\newblock in: \bibinfo{booktitle}{IEEE Int. Conf. on Acoustics, Speech and
  Signal Process.}, \bibinfo{year}{2017}, pp. \bibinfo{pages}{4104--4108}.
\bibitem[{Zhu et~al.(2020)Zhu, Iglesias, Marques, and
  Segarra}]{zhu_2020_estimating}
\bibinfo{author}{Y.~Zhu}, \bibinfo{author}{F.~J. Iglesias},
  \bibinfo{author}{A.~G. Marques}, \bibinfo{author}{S.~Segarra},
\newblock \bibinfo{title}{Estimating network processes via blind identification
  of multiple graph filters},
\newblock \bibinfo{journal}{IEEE Trans. Signal Process.} \bibinfo{volume}{68}
  (\bibinfo{year}{2020}) \bibinfo{pages}{3049--3063}.
\bibitem[{Fazel et~al.(2001)Fazel, Hindi, and Boyd}]{Fazel2001}
\bibinfo{author}{M.~Fazel}, \bibinfo{author}{H.~Hindi}, \bibinfo{author}{S.~P.
  Boyd},
\newblock \bibinfo{title}{A rank minimization heuristic with application to
  minimum order system approximation},
\newblock in: \bibinfo{booktitle}{American Control Conf.},
  \bibinfo{year}{2001}, pp. \bibinfo{pages}{4734--4739}.
\bibitem[{Tropp(2006)}]{Tropp2006}
\bibinfo{author}{J.~A. Tropp},
\newblock \bibinfo{title}{Just relax: {C}onvex programming methods for
  identifying sparse signals in noise},
\newblock \bibinfo{journal}{IEEE Trans. Inf. Theory} \bibinfo{volume}{52}
  (\bibinfo{year}{2006}) \bibinfo{pages}{1030--1051}.
\bibitem[{Fazel et~al.(2003)Fazel, Hindi, and Boyd}]{Fazel2003}
\bibinfo{author}{M.~Fazel}, \bibinfo{author}{H.~Hindi}, \bibinfo{author}{S.~P.
  Boyd},
\newblock \bibinfo{title}{Log-det heuristic for matrix rank minimization with
  applications to {H}ankel and {E}uclidean distance matrices},
\newblock in: \bibinfo{booktitle}{American Control Conf.},
  \bibinfo{year}{2003}.
\bibitem[{Cour et~al.(2006)Cour, Srinivasan, and Shi}]{Cour2006}
\bibinfo{author}{T.~Cour}, \bibinfo{author}{P.~Srinivasan},
  \bibinfo{author}{J.~Shi},
\newblock \bibinfo{title}{Balanced graph matching},
\newblock in: \bibinfo{booktitle}{Neural Inf. Process. Systems},
  \bibinfo{year}{2006}.
\bibitem[{Bollob{\'a}s(1998)}]{bollobas1998random}
\bibinfo{author}{B.~Bollob{\'a}s}, \bibinfo{title}{Random Graphs},
  \bibinfo{publisher}{Springer}, \bibinfo{year}{1998}.

\end{thebibliography}

\end{document}